\documentclass[a4paper,fleqn,usenatbib]{mnras}

\usepackage{newtxtext,newtxmath}

\usepackage[T1]{fontenc}
\usepackage{ae,aecompl}

\usepackage{graphicx}	
\usepackage{amsmath}	
\usepackage{hyperref}

\usepackage[dvipsnames]{xcolor} 
\usepackage{subfig}
\usepackage[export]{adjustbox}
\usepackage{dcolumn}
\providecommand{\mc}[1]{\multicolumn{1}{c}{#1}}

\providecommand{\pkg[1]}{\textup{\textsc{#1}}}

\providecommand{\e}[1]{\times 10^{#1}}

\title[Faint carbon emission in high-$z$ radio galaxies ]{Faint [\ion{C}{I}](1-0) emission in $z \sim 3.5$ radio galaxies}

\author[S. Kolwa et al.]{S. Kolwa,$^{1,2}$\thanks{E-mail: sthabile@idia.ac.za} 
C. De Breuck,$^{3}$
J. Vernet,$^{3}$
D. Wylezalek,$^{7}$
W. Wang,$^{7}$
G. Popping,$^{3}$
\newauthor
A.W.S. Man,$^{4,5}$
C.M. Harrison,$^{6}$
P. Andreani,$^{3}$
\\
$^{1}$Physics Department, University of Johannesburg, 5 Kingsway Ave, Rossmore, Johannesburg, 2092, South Africa\\
$^{2}$Inter-University Institute for Data Intensive Astronomy, Department of Astronomy, University of Cape Town, Rondebosch 7701, South Africa\\
$^{3}$European Southern Observatory, Karl-Schwarzschild-Str. 2 85748 Garching bei M\"unchen, Germany \\
$^{4}$Dunlap Institute for Astronomy and Astrophysics, University of Toronto, 50 St George Street, Toronto ON, M5S 3H4, Canada\\
$^{5}$Department of Physics \& Astronomy, The University of British Columbia, 6224 Agricultural Road, Vancouver, BC V6T 1Z1, Canada\\
$^{6}$School of Mathematics, Statistics and Physics, Newcastle University, Newcastle upon Tyne, NE1 7RU, UK\\
$^{7}$Astronomisches Rechen-Institut, Zentrum f\"ur Astronomie der Universit\"at Heidelberg, M\"onchhofstra$\beta$e 12-14, 69120 Heidelberg, Germany
}

\date{Accepted XXX. Received YYY; in original form ZZZ}

\pubyear{2022}

\begin{document}
\label{firstpage}
\pagerange{\pageref{firstpage}--\pageref{lastpage}}
\maketitle

\begin{abstract}
We present Atacama Large Millimeter/sub-millimeter Array (ALMA) neutral carbon, [\ion{C}{I}](1-0), line observations that probe molecular hydrogen gas (H$_2$) within seven radio galaxies at $z=2.9 - 4.5$ surrounded by extended ($\gtrsim$100 kpc) \Lya nebulae. We extract [\ion{C}{I}](1-0) emission from the radio-active galactic nuclei (AGN) host galaxies whose positions are set by near-infrared detections and radio detections of the cores. Additionally, we place constraints on the galaxies' systemic redshifts via \ion{He}{II} $\lambda1640$ lines seen with the Multi-Unit Spectroscopic Explorer (MUSE). We detect faint [\ion{C}{I}] emission in four out of seven sources. In two of these galaxies, we discover narrow line emission of full width at half maximum $\lesssim100$ km s$^{-1}$ which may trace emission from bright kpc-scale gas clouds within the ISM. In the other two [\ion{C}{I}]-detected galaxies, line dispersions range from $\sim100 - 600$ km s$^{-1}$ and may be tracing the rotational component of the cold gas. Overall, the [\ion{C}{I}] line luminosities correspond to H$_2$ masses of $M_{\rm H_2,[\ion{C}{I}]} \simeq (0.5-3) \times 10^{10}$ M$_\odot$ for the detections and $M_{\rm H_2,[\ion{C}{I}]}<0.65\e{10}$ M$_\odot$ for the [\ion{C}{I}] non-detections in three out of seven galaxies within the sample. The molecular gas masses in our sample are relatively low in comparison to previously reported measures for similar galaxies which are $M_{\rm H_2,[\ion{C}{I}]} \simeq (3-4) \times 10^{10}$ M$_\odot.$ Our results imply that the observed faintness in carbon emission is representative of a decline in molecular gas supply from previous star-formation epochs and/or a displacement of molecular gas from the ISM due to jet-powered outflows.
\end{abstract}



\section{Introduction}\label{section:introduction}
Observations of cold gas in radio galaxies (HzRGs) which host high-power jetted active galactic nuclei (AGN) are an excellent tool for studying the impact of radio-mode feedback on baryonic matter within the interstellar medium and extended halo of a high-mass galaxy. The distinguishing features of radio galaxies at high-redsift ($z > 1$) are their high-luminosity nuclear emission, prominent radio-jets, and high stellar masses which typically range from $M_\star \simeq 10^{11} - 10^{12}$ M$_\odot$ \citep{Seymour2007,deBreuck2010}. Such a combination of physical features make HzRGs particularly important sites for examining the co-evolution of galaxies and their central supermassive black-holes as well as feedback mechanisms regulated by the active galactic nucleus (AGN).

Within the past three decades, it has become increasingly apparent that HzRGs are also commonly associated with enormous \Lya nebulae (ELANe) that can extend out to distances of $\gtrsim100$ pkpc from their nuclei. The combination of powerful radio jets and enormous haloes of ionised gas surrounding the galaxy mean that HzRGs are also very convenient probes for investigating the interplay between ionised gas and radio emission \citep{McCarthy1993,Reuland2003,Villar-martin2003,Villar-martin2007,Humphrey2007,Swinbank2015,Morais2017,Silva2018b}. ELAN surrounding HzRGs represent the warm ionised gas component ($\rm T \sim10^4 - 10^5$ K) of the circumgalactic medium (CGM) where turbulent gas motions is observed up to tens of kpc from the AGN while on the outermost envelopes of the halo ($\gtrsim 10$ pkpc) warm gas tends to be more kinematically quiet \citep{Villar-martin2003}.

Generally, the CGM of a galaxy is defined as the gas region through which baryonic matter is transported (via accretion, and outflows) between the intergalactic medium (IGM) and the interstellar medium (ISM). With the CGM being multi-phase, consisting of cold molecular, neutral as well as ionised gas, it has become an important subject for investigation in galaxy evolution, allowing us to determine how multi-phase gas components interact physically with one another throughout the extent of a galaxy's halo \citep{Steidel2010,Tumlinson2017}.

The ionised component of the CGM surrounding galaxies that host AGN has been examined in numerous studies involving observations, photoionisation modelling and theoretical simulations. These studies have made significant use of integral field unit (IFU) spectrographs to observe the ionised gas, providing a more detailed look into the kinematics of the \Lya nebulae surrounding galaxies which had been discovered a decade or two prior to the advent of IFUs. The 3D imaging capability of IFUs provide a crucial in-depth view of the gaseous environments of galaxies up to 100s of kpc from their nuclei. \Lya line emission traced directly from the target galaxy with an IFU is equally useful in studying the warm gas component of the CGM \citep{Borisova2016,Ginolfi2018,Arrigoni-Battaia2018,Arrigoni-Battaia2019,Fossati2021}. When the CGM of a galaxy coincides with a quasar sight-line, absorption-line measures in the quasar spectrum become a very useful diagnostic for the chemical composition and kinematics of the foreground galaxy's CGM \citep{Bielby2017,Peroux2017,Dutta2020}. 

For HzRGs in particular, instruments such as the Spectrograph for INtegral Field Observations in the INfrared \citep[SINFONI; e.g.][]{Nesvadba2006b,Nesvadba2017} and the Multi-Unit Spectroscopic Explorer \citep[MUSE; e.g.][]{Swinbank2015,Gullberg2016b,Vernet2017,Kolwa2019,Falkendal2021,Wang2021} have illustrated the impact of the powerful radio jets on the kinematics of emitting gas, in terms of producing outflows and increasing gas turbulence, showing that jets can drive out molecular gas and effectively shut down the fuel supply for star-formation.

While all of the above observations have been beneficial in providing information on the properties of ionised gas in radio-loud AGN hosts, a systematic CO survey to trace H$_2$ in a wide sample of HzRGs has yet to be performed. This is the case despite the fact that low-$J$ transitions of CO have minimal excitation requirements (i.e., low critical densities and $E/{\rm k_B}$ values). This fact implies that CO lines can easily trace H$_2$ clouds irrespective of their thermal states at low gas densities $n \sim$ $10^2$ cm$^{-3}$ which are typically seen in optically thick CO(1-0) gas. While CO has clear advantages as a molecular gas tracer, solely relying on it comes with drawbacks. For instance, in jetted AGN, CO within the ISM can be photodissociated by cosmic rays, effectively destroying CO molecules and reducing the CO/H$_2$ abundance in giant molecular clouds (GMCs) \citep{Bisbas2015,Bisbas2017}. 

Within this framework, neutral carbon fine-structure line transitions, [\ion{C}{i}] $^3P_1 \rightarrow$ $^3P_0$ and [\ion{C}{i}] $^3P_2 \rightarrow$ $^3P_1,$ hereinafter [\ion{C}{i}](1-0) and [\ion{C}{i}](2-1), have been introduced as an alternative tracer for H$_2$ in GMCs. Selecting [\ion{C}{i}] as a tracer may be especially wise given that radio galaxies have jetted AGN that produce high-energy cosmic rays which are capable of photodissociating CO in molecular gas. As a result, [\ion{C}{i}] lines may retain their ability to trace H$_2$ in clouds irradiated by cosmic rays (CRs). This situation may change as one shifts away from the nuclear region and towards the interstellar medium (ISM) where the number density of ionising photons is relatively lower than it is in a galaxy's nuclear region \citep{PapadopoulosThiViti2004}
 
Previously, a pilot study of molecular gas in ultra-luminous infrared galaxies (ULIRGs) found reasonable agreement between the H$_2$ masses inferred from [\ion{C}{i}] and low-$J$ CO lines \citep[e.g.][]{PapadopoulosGreve2004}. While more recently, observations in \citet{Papadopoulos2018} have shown that high energy cosmic rays (CRs) produced by star-formation and active galactic nuclei (AGN) can provide the right conditions for creating a [\ion{C}{I}]-rich/CO-poor molecular gas phase. Another advantage to using [\ion{C}{i}] line tracers is that they provide a straightforward interpretation of the molecular clouds allowing us to infer $\rm H_2$ mass. This is due to [\ion{C}{i}] lines having typically low optical depths in comparison to CO lines. Hence, the [\ion{C}{i}] line surface brightness correlates more directly with the mass of neutral carbon and also the $\rm H_2$ mass of a molecular cloud \citep{Nesvadba2019}. Furthermore, both [\ion{C}{i}](1-0) and [\ion{C}{i}](2-1) transitions have sufficiently low critical densities of $n_{10} = 500$ cm$^{-3}$ and $n_{21} = 10^3$ cm$^{-3},$ respectively.

One of the first major high-redshift ($z > 2$) [\ion{C}{i}] line surveys was carried out by \citet{Walter2011} and showed emission in sub-millimetre galaxies and quasar hosts. Since then, several [\ion{C}{i}] surveys have been performed with the goal of tracing H$_2$ and inferring molecular gas masses and the dynamics thereof for both lensed and unlensed galaxies at high-$z$ \citep{Alaghband-Zadeh2013,Bothwell2017,Valentino2018,Nesvadba2019}. Adding to these surveys are several single-source studies which reveal the presence of [\ion{C}{i}] emission in star-forming galaxies \citep{Popping2017,Andreani2018}. For HzRGs, [\ion{C}{i}](2-1) has traced H$_2$ in PKS 0529-549 at $z=2.57$ \citep{Lelli2018,Man2019}. Within the halo of the Spiderweb Galaxy (MRC 1138-262 at $z\simeq2.16$), \citet{Gullberg2016b} have reported broad [\ion{C}{i}](2-1) emission at a projected distance of $d\simeq4$ kpc from the main galaxy's radio core. In addition to this, [\ion{C}{i}](1-0) emission, within the Spiderweb's gas halo, has been detected between $d \simeq 17$ and 70 kpc from its core \citep{Emonts2018}. Other recent studies of [\ion{C}{i}](1-0) line emission in HzRGs have been carried out for two sources: 4C+14.17 at $z=3.8$ \citep{Nesvadba2020} and 4C+19.71 at $z\simeq3.6$ \citep{Falkendal2021}. With all these results, it is clear that the drive to use [\ion{C}{i}] as a molecular gas tracer has certainly caught on thus strengthening its case for being a similarly and perhaps even more reliable H$_2$ tracer than CO \citep[e.g.][]{Dunne2022}.

For tracing [\ion{C}{i}](1-0) with sub-arcsecond resolution, the Atacama Large Millimeter-submillimeter Array (ALMA) has frequency bands 3 and 4 which are best suited for probing emission from this atomic transition at a redshift range of $z=2.9 - 4.5$ where rest-UV lines are detectable within the observational window of the Multi-Unit Spectroscopic Explorer (MUSE). Hence, with a combined ALMA and MUSE dataset, we can perform a detailed kinematic analysis of both the ionised and molecular gas. In doing so, we are capable of assessing the probable effects of shock-induced turbulence from radio jets as well as photon-heating on the molecular gas within the ISM \citep{Mahony2013,Morganti2018} and possibly also the CGM \citep{McNamara2016}.
 
We have therefore constructed a sample of seven HzRGs with redshifts spanning $z=2.9 - 4.5.$ This galaxy sample has been observed with both ALMA in [\ion{C}{i}](1-0) and MUSE which probes the \Lya emission. In tandem, the ALMA and MUSE observations provide us with a multi-wavelength view of the cold ($\sim$ 10 K) and warm ($\sim10^4 - 10^5$ K) gas phases of baryonic haloes surrounding radio galaxies. With such an imaging set, we are capable of gauging the impact of high-powered radio jets on the multi-phase gas within both a galaxies' ISM and its surrounding CGM.

This paper is structured in the following way: Section 2 describes the sample selection; Section 3 provides an overview of the data acquisition and analysis procedures; Section 4 explains the method for the analysis; and in Section 5, we describe the overall results of this study and explore the implications of our findings. A summary is provided in Section 6. Throughout, we have made use of $\Lambda$CDM cosmology as defined by \citet{Planck2016} where $\rm H_0 = 67.8~km~s^{-1}~Mpc^{-1}$ and $\rm \Omega_m = 0.308.$  


\section{Galaxy Sample Selection}\label{section:sample-selection}
The radio galaxies covered in this work are drawn from legacy radio surveys. Our sample, in particular, comprises seven targets which are listed sources in the Molongolo Reference Catalogue \citep[MRC;][]{Large1981} and Fourth Cambridge Catalogue \citep[4C;][]{PilkingtonScott1965}. The targets are radio sources with optical counterparts i.e. radio galaxies with rest-frame 5 GHz luminosities of $\geq10^{25}$ W Hz$^{-1}$ which imply the presence of radio-loud AGN \citep{Miller1990}. Currently, radio galaxies are known out to redshifts of $z$=5.72 \citep{Saxena2018}. Generally speaking, those above a redshift of 2 (i.e., $z > 2$) are referred to as high-$z$ radio galaxies (HzRGs) which are rare and require a significant amount of optical and infrared spectroscopic observing time for reliable constraints on their redshifts to be made. Despite this, several radio selection techniques have already been applied to pre-select high-$z$ radio galaxy candidates \citep[e.g.][]{Drouart2020}. As a result of the stringency of pre-selections, however, the space density of HzRGs is not sufficiently constrained, at the present time.

Our ALMA+MUSE sample has been selected from a large survey program designed to study stellar mass build up and star-formation in distant radio galaxies at $1.0 < z < 5.2$ with the {\it Spitzer} Space Telescope \citep{Seymour2007,deBreuck2010}. This dedicated {\it Spitzer} HzRGs program consists of 71 HzRGs which have stellar masses which range from $M_\star \simeq10^{11} - 10^{11.5}$ M$_\odot$ (details on stellar mass inferences are given in Section \ref{subsection:ancillary-data}). The $3.6-850~\mu$m {\it Spitzer} and {\it Herschel} photometry first reported by \citet{Drouart2014} have been combined with ALMA continuum observations to obtain SED fits that provide the star-formation rate measures for the galaxies in our sample \citep{Falkendal2019}. 

With the goal of simultaneously tracing the ionised and molecular gas components of the extended halos of HzRGs, we have defined a sub-sample of seven HzRGs that are observable both in \Lya with MUSE and in [\ion{C}{i}] with ALMA. For ALMA, we targeted the neutral carbon, ground-state transition line [\ion{C}{i}] $^3P_1 \rightarrow$ $^3P_0$ (hereinafter [\ion{C}{i}](1-0) at $\nu_{\rm rest}$ = 492.161 GHz), with simultaneous coverage of $^{13}$CO(4-3) at $\nu_{\rm rest}$ = 440.765 GHz. This was done with the aim of tracing the cold gas component within the host galaxies and their extended haloes. At the $2.9 \lesssim z \lesssim 4.5$ redshift coverage of our sub-sample, the [\ion{C}{i}](1-0) line can be observed in ALMA bands 3 and 4. 

On the other hand, the MUSE survey targeted rest-frame UV emission lines that are observable within the spectral window of MUSE (in the wide-field mode) which covers a wavelength range of $\lambda_{\rm obs}=4800-9300~\ang.$ The MUSE observations are ideal for tracing warm ionised gas at redshifts of $z=2.9 - 6.6$ where the \Lya$\lambda1216~\ang$ line falls within the MUSE spectral window. The blue edge of MUSE sets the lower redshift range of our subsample.

\begin{figure*}
 \includegraphics[width=0.75\textwidth]{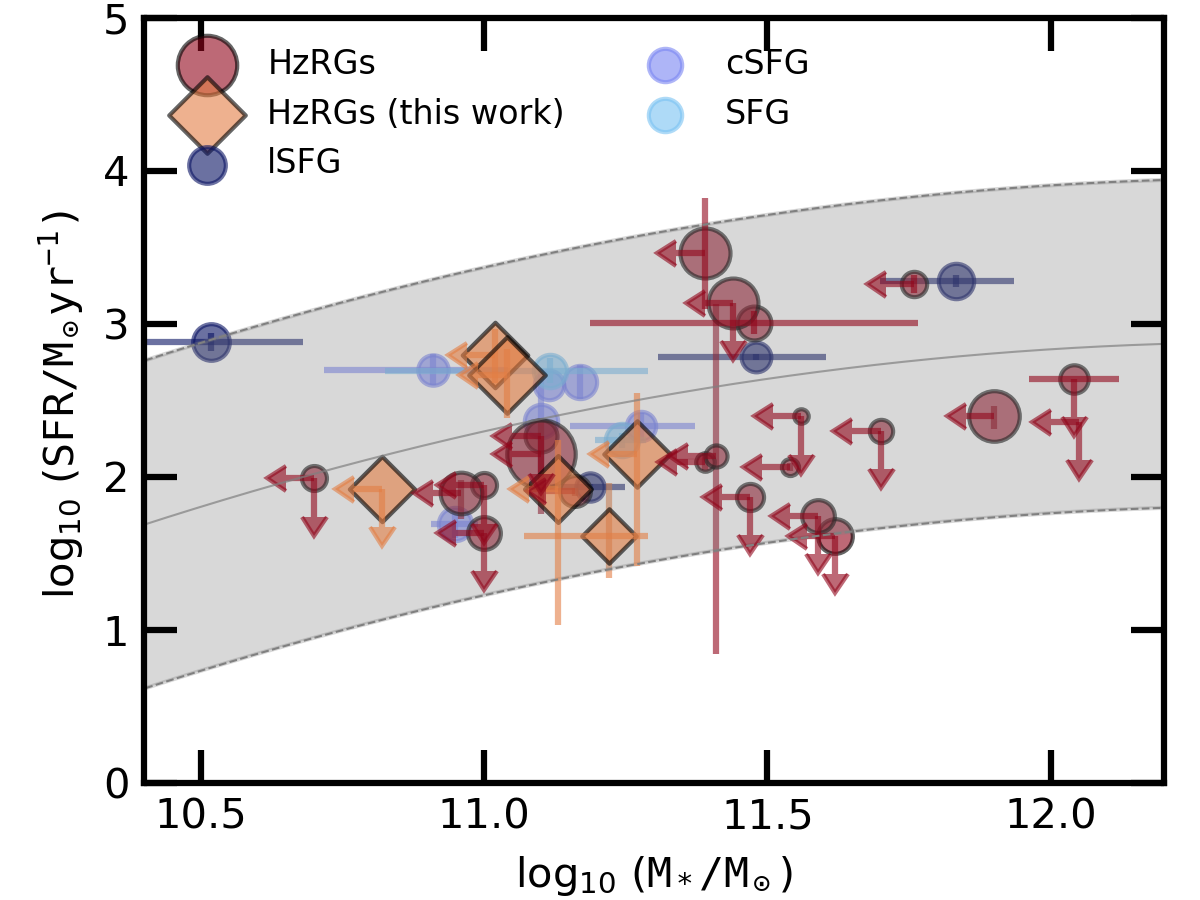}
 \caption{The star-formation rate as a function of stellar mass for HzRGs and several classes of SFGs. The HzRGs from the ALMA+MUSE sample (orange) presented in this work are shown alongside an HzRG sample \citep{Falkendal2019} in red. Lensed SFGs (lSFGs in navy blue), compact SFGs (blue) and normal SFGs (sky blue) are shown. The main sequence of star-forming galaxies from \citet{Schreiber2015} at $z=2.52,$ the median redshift across all the galaxies depicted, is shown in grey with a 0.3 dex region of scatter in SFR which corresponds to a $1\sigma$ dispersion in $\log_{10} \rm (SFR / M_\odot yr^{-1})$ which is based on empirical results from a flux-limited sample of galaxies with {\it Spitzer} MIPS \citep{Noeske2007,Elbaz2007}. The data point sizes are scaled as $10z^2$ where $z$ is the redshift of the galaxy. Literature references of the galaxy samples are provided in the text. We include a sample of a {\it Herschel}-detected, lensed SFGs using their magnification-corrected SFR and $M_\star$ measures \citep{Sharon2013,Bothwell2013b,Dessauges-Zavadsky2015,Nayyeri2017}. The figure also shows a sample of six compact SFGs \citep{Tadaki2015,Spilker2016,Popping2017,Tadaki2017b,Barro2017}. Additionally, a sample of normal SFGs have been adapted from \citet{Tadaki2015} and are shown in the figure as well.}
 \label{fig:SFR-M-star}
\end{figure*}

To place our sample into the general context of high-$z$ galaxies, we construct a SFR-$M_\star$ plane (Fig. \ref{fig:SFR-M-star}) where our targets are shown against HzRGs from previous studies and star-forming galaxies (SFGs) at $z > 1.0.$ The SFRs for the HzRGs represented in the figure have been adapted from \citet{Falkendal2019} who make use of {\it Herschel}, {\it Spitzer} and ALMA photometry to perform SED-template fitting and derive the infrared (IR) luminosities required to estimate SFRs for the galaxies. 
Additionally, \citet{Seymour2007} use {\it Spitzer} photometry to perform SED-fits that yield $M_\star$ as upper limits for the 26 out of 29 HzRGs from our study as well as the literature in Fig. \ref{fig:SFR-M-star}. $M_\star$ constraints are made for 3 out of the 29 HzRGs shown.

\section{Observations and Data Reduction}
  \subsection{ALMA}\label{section:alma-details}
The radio galaxies in our sample were observed in ALMA bands 3 \citep{Kerr2014} and 4 \citep{Asayama2014} during ALMA Cycle 3 under the project ID 2015.1.00530.S (PI: De Breuck) on the dates provided in Table \ref{table:hzrgs-obs-alma-muse}. The correlator configurations was set to observe two contiguous spectral windows covering the [\ion{C}{i}](1-0) line and continuum in one side-band and the remaining two spectral windows simultaneously covering the $^{13}$CO(4-3) line and continuum in the other side-band. All four spectral windows were recorded in Frequency Domain Mode with a bandwidth 1875 MHz and spectral resolution of 3.904 MHz. 

We generated the calibrated measurement sets using the calibration scripts provided by the ALMA Observatory using \pkg{casa} (\textit{Common Astronomy Software Applications}) versions 4.5.1, 4.5.3, 4.6.0, and 4.7.2 \citep{mcmullin2007}. For the imaging, we used the data reduction pipeline (\pkg{casa}-6.2.1.7 and a natural {\it uv}-weighting ({\tt robust} parameter 2) to optimise the signal-to-noise ratio (S/N) at the expense of spatial resolution. As the CGM emission can be quite spatially extended \citep[e.g.][]{Emonts2018}, we also made an attempt at tapering the 12m {\it uv}-data with 2\arcsec, 3\arcsec, and 4\arcsec~Gaussian width options. For all sources, with the exception of MRC 0943-292, an increase in noise due to the reduced amount of data decreased the S/N data overall. Hence, we opted to continue forward with the untapered images for which the spatial resolution is provided in Table \ref{section:alma-details}. Continuum subtraction was performed using the {\it uvcontsub()} task and spectral-line cubes were generated with a range of velocity resolutions from 9 to 64 km s$^{-1}$, where we selected the binning that provides the most optimal line S/N for the observed line width. The images were primary beam corrected with the \pkg{casa} task {\it impbcor()}. For the 1D spectra, pixels were averaged over 2\arcsec~diameter apertures. The moment-0 maps (i.e. narrow-band images) were created with \pkg{casa} {\it immoments()} in which we images were integrated over frequency ranges shown in Fig.~\ref{fig:galaxy-spectra-images}. The $1\sigma$ rms levels in the [\ion{C}{I}] spectra are two orders of magnitude smaller than the [\ion{C}{I}] contour levels shown in the corresponding images due to the velocity integration performed during the moment-0 map creation. 

\begin{table*}
  \centering
  \caption[ALMA observations of HzRGs]{ALMA and MUSE observation details for the $2.9 \lesssim z \lesssim 4.5$ radio galaxy sample. Columns (1) and (2) show the galaxy catalogue names and literature redshifts. Column (3) indicates the start and end dates for the completion of the 12m ALMA observations, column (4) shows the on-source integration time ($t_{\rm int.}$). In column (5), we provide the full-width at half maximum (FWHM) of the synthesised beam along its major and minor axes. Column (6) shows the MUSE observing dates. Column (7) indicates the source exposure time ($t_{\rm exp.}$). Column (8) indicates the seeing conditions during observations which are obtained from the average FWHM of the brightest foreground star in the field.}
  \label{table:hzrgs-obs-alma-muse}
  \begin{tabular}{| l | l l l | l l l |} 
  \hline
  & {\bf ALMA} & & & {\bf MUSE} & & \\
 & & & & & & \\
  Galaxy & Observing dates & $t_{\rm int}$  & bmaj $\times$ bmin & Observing Dates & $t_{\rm exp.}$ & Seeing \\
  &      & (min)               & (arcsec$^2$)&   & (min) & (arcsec)   \\
 & & & & & & \\
  \hline 
  & & & & & & \\
  MRC 0943-242  & 04/03/2016 - 20/05/2016 	& 45  & $2.09\times1.33$  & 21/02/2014 - 18/01/2016  & 312 & 0.65 \\
  TN J0205+2242 & 08/03/2016 - 13/09/2016 	& 44 	& $2.28 \times 1.81$  & 03/12/2015 - 08/12/2015  & 254 & 0.73 \\  
  TN J0121+1320 & 06/03/2016 - 24/09/2016 	& 40 	& $1.98 \times 1.47$  & 06/10/2015 - 28/08/2016  & 318 & 0.83 \\
  4C+03.24 & 06/03/2016   				& 38  & $2.23 \times 1.62$  & 17/06/2017 - 18/06/2017  & 75 & 0.63 \\
  4C+19.71 & 06/03/2016 - 24/09/2016 	& 39  & $1.98 \times 1.81$ & 08/06/2016 - 02/09/2016  & 350 & 1.03 \\
  TN J1338-1942 & 16/03/2016 - 22/09/2016 	& 40 	& $1.98 \times 1.61$  & 30/04/2014 - 30/06/2014  & 535 & 0.77 \\
  4C+04.11 & 05/03/2016 - 03/05/2016 	& 43 	& $2.24 \times 2.17$  & 03/12/2015 - 15/12/2015  & 254 & 0.88 \\
  & & & & & & \\
  \hline
  \end{tabular}
\end{table*}

  \subsection{MUSE}
From the MUSE IFU spectrograph, we acquired rest-UV line and continuum observations for the sample of radio galaxies. Mounted on VLT Yepun (UT4), MUSE covers a spectral range of $4800-9300~\ang$ where the spectral resolution ranges from $2.82-2.74~\ang$ between the blue and red wavelength extremities of the MUSE spectral window. All of the galaxies, excluding for 4C+03.24 and MRC 0943-242, were observed in WFM-NOAO-N (nominal wide-field mode with no adapative optics system in place) mostly under the program IDs 096.B-0752 and 097.B-0323 (PI: Vernet) on several nights between 2015 October and 2016 September (see Table \ref{table:hzrgs-obs-alma-muse}). 

The galaxy, 4C+03.24, was observed under program IDs 060.A-9100(G) during the GALACSI/WFM commissioning run in 2017 June. TN J1338-1942 was observed under 60.A-9318(A) and 060.A-9100(B) during the science verification and second MUSE commissioning run, from 2014 April to June. For MRC 0943-242 which was observed in WFM-NOAO-E (extended wide-field mode, no adapative optics), we also include the data of obtained during first commissioning in 2014 February under 60.A-9100(A).

We processed the raw observations using a standard data reduction procedure in \pkg{esorex} v.2.8.4 \citep{Weilbacher2020} which produced the final data-cube for each galaxy. We then ran a principal component analysis based procedure called {\it Zurich Atmosphere Purge} (\pkg{zap}~2.0) \citep{Soto2016} on the reduced data-cubes in order to subtract telluric sky line emission where it is most present at the red end of the MUSE spectral window. 

To create the \Lya narrow-band images, the MUSE data-cubes were integrated along the spectral axes over $15~\ang$ around the peak in \Lya emission in each source. The images have a sampling rate of $0.2 \times 0.2$ arcsec$^2$ per pixel which is seeing-limited to an average $1.0 \arcsec$ in our observations. We note that some of the MUSE data on individual sources has already been published in previous papers \citep{Swinbank2015,Gullberg2016a,Vernet2017,Falkendal2019,Kolwa2019,Wang2021}. Additionally, detailed tomography of the \Lya nebulae for eight radio-loud AGN (including the seven galaxies from our sample) are reported in \citet{Wang2023}.  

  \subsection{Ancillary data}\label{subsection:ancillary-data}
Optical, near-infrared (NIR) and radio data are available for the HzRG sample which we have selected. The multi-wavelength dataset includes {\it Hubble Space Telescope} (HST) wide-band imaging obtained with Wide-field Planetary Camera 2 (WFPC2) with the F702W filter which has a spectral coverage of $5865.66-8433.19~\ang$ useful for tracing the continua of O \& B-type stars in high-$z$ sources \citep{Pentericci1997}. The reduced HST images images are accessible via the \href{https://hla.stsci.edu/hlaview.html}{Hubble Legacy Archive (HLA).} 

The {\it Spitzer} Space Telescope images of radio galaxies reveal emission from evolved stellar populations \citep{Seymour2007}. We have obtained rest-frame $K-$band detections from the Infrared Array Camera (IRAC; $3.6-8.0~\mu$m). In addition to those taken by the other {\it Spitzer} instruments: Infrared Spectrograph (IRS; 16 $\mu$m) and Multiband Imaging Photometer for {\it Spitzer} (MIPS; $24-160~\mu$m), {\it Spitzer} images we use in this study are available and accessible from a dedicated \href{http://www.eso.org/~cdebreuc/shzrg/}{{\it Spitzer} HzRGs (SHzRGs)} archive. The near-IR, {\it Spitzer} and {\it Herschel} data \citep{deBreuck2010,Drouart2014} provide stellar masses and star formation rates (or limits thereof) for our entire sample, which allows us to derive the evolutionary state of the galaxies relative to the Main Sequence of star-forming galaxies (see Fig.~\ref{fig:SFR-M-star}). We note that the stellar masses reported by \citet{Seymour2007} and \cite{deBreuck2010} are often upper limits due to potential contributions from the hot dust torus emission to the rest-frame 1.6\,$\mu$m photometry. However, this approach is conservative, as most of the stellar populations are consistent with the $K$-band photometry. As a result, the real stellar masses in Fig.~\ref{fig:SFR-M-star} are expected to be sufficiently close to the upper limits, which unlike normaly upper limits, have been measured from significantly deep photometry. For the radio imaging of these sources, we obtained archival C-band (4.8 GHz) and X-band (8.3 GHz) images from the Karl G. Jansky Very Large Array (VLA) that indicate the locations of the radio hotspots \citep{Carilli1997,Pentericci2000}. 

\section{Data Analysis}
  \subsection{Systemic Redshift Estimation}\label{section:HeII-fits}
The galaxies in our combined ALMA+MUSE sample have spectroscopic redshifts from the literature which are based on a variety of line tracers which we provide references of in table \ref{table:redshifts}. With the advent of optical 3D data cubes, the complex kinematic structure of extended ionised gas haloes has become more apparent revealing outflowing ionised gas from AGN host galaxies \citep[e.g.][]{Molyneux2019,Couto2020,Riffel2023}. The presence of such components can lead to offsets of several hundreds of km s$^{-1}$ when estimating systemic redshifts, even when non-resonant lines such as \ion{He}{II} are used \citep[e.g.][]{Kolwa2019}. 

The \ion{He}{II} $\lambda$1640 recombination line is our benchmark for setting the systemic redshift of each galaxy. The \ion{He}{II} profiles are extracted from the galaxy positions in the MUSE data-cubes for all the sources except 4C+03.24 which does not have detected \ion{He}{II} emission at its core, and 4C+19.71, where the \ion{He}{II} line coincides with a spectral region affected by skyline residuals. The 1D spectra are extracted from aperture diameters of 5 pixels or $1\arcsec$ at the positions of the host galaxies which are determined from the radio cores \citep{Carilli1997,Pentericci2000} and the {\it Spitzer} imaging which indicates the extent of the evolved stellar distribution (see Fig \ref{fig:galaxy-spectra-images}). The size of the aperture is chosen (i) to match the seeing element (of width $0.9\arcsec\sim1.0\arcsec$) estimated from stars in the MUSE fields of view; (ii) to maximise the spectral S/N ratio; (iii) to avoid contamination from non-systemic line emission components further away from the core which may emerge within the line profile as the aperture size is increased.

\begin{figure}
 \centering
\subfloat[]{\includegraphics[width=0.75\columnwidth]{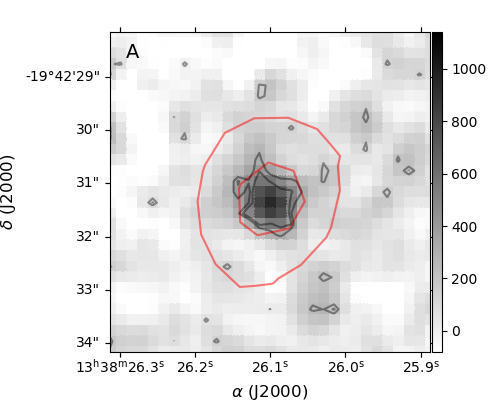}}\\
\subfloat[]{\includegraphics[width=0.75\columnwidth]{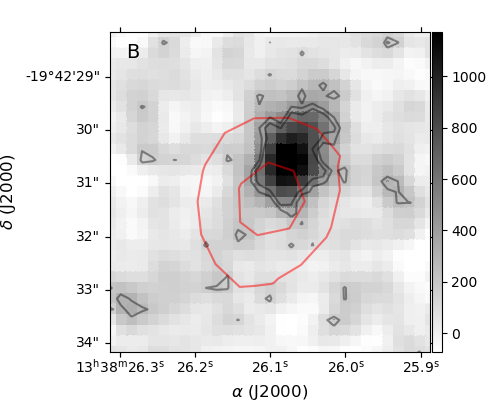}}
\caption{MUSE-detected \ion{He}{II} $\lambda1640$ emission in TN J1338-1952 integrated over $8345.65 - 8362.38~\ang$ in panel A and $8376 - 8393~\ang$ in panel B (grey-scale with black contours) against the {\it Spitzer} IRAC2 continuum (red).}
\label{fig:HeII-IRAC2-TNJ1338}
\end{figure}

We began by applying a Gaussian fit to the \ion{He}{II} profiles of MRC 0943-242, TN J0205+2242, TN J0121+1320 and 4C+04.11 where the Gaussian peak ($A$), width ($\sigma$) and centre ($\lambda_c$) were fit to the data with $A$ as a free parameter and sigma constrained within the wavelength range $2-15~\ang$ and $\lambda_c$ is allowed to deviate $\pm5~\ang$ from its initial value. The \ion{He}{II} emission of TN J1338-1942 is spatially extended and, when extracted to form a 1D spectrum, it is seen to have two distinct spectral profiles: one at the position of the host galaxy and the other offset in projected distance from the host galaxy illustrated in Fig. \ref{fig:HeII-IRAC2-TNJ1338}. As a result, a double Gaussian profile, with components that are independent of one another, is fit to the spectrum. The peak, width and line centres are constrained by the same boundary conditions as those used for the single Gaussian fits. 

In the case of a single emission component, the Gaussian centre is used to infer the systemic redshift of the source. For a double emission component, we needed to carefully select which Gaussian component represents the systemic redshift. We therefore look for a spatial overlap between the \ion{He}{II} and $K-$band continuum from {\it Spitzer}/IRAC2 to confirm which emission component in \ion{He}{II} traces the host galaxy and should thus represent the systemic redshift.

As a result, we obtain 1D spectral line models which are shown in Fig. \ref{fig:galaxy-spectra-images}, with the results being listed in Table~\ref{table:redshifts}. Where [\ion{C}{i}] is detected, the \ion{He}{II} and [\ion{C}{i}] redshifts are fully consistent within the margin of error. Given the low S/N of our [\ion{C}{i}] data, the uncertainties of the [\ion{C}{i}] redshifts are higher than on the \ion{He}{II} redshifts (Table~\ref{table:redshifts}) hence the current ALMA data may not be able to improve the overall systemic redshift accuracy but they do allow for a consistency check. In two sources, the \ion{He}{II} line cannot be fit due to its low S/N. Here, the [\ion{C}{i}] line is used to constrain the systemic redshift for the 4C+03.24 and 4C+19.71. In 4C+03.24, \ion{He}{II} is not detected and we fix the redshift to the centre of the [\ion{C}{i}] line (bottom panel of Fig. \ref{fig:galaxy-spectra-images}). In 4C+19.71, \ion{He}{II} is strongly affected by skyline residuals \citep[e.g][]{Falkendal2021}, even though some of the line emission is discernible, we provide as a Gaussian profile that is fixed to the [\ion{C}{i}] redshift (top panel of Fig. \ref{fig:galaxy-spectra-images}) and label this {\it \ion{He}{II} fix} in the figure.

\begin{table*}
  \centering
  \caption[]{Redshifts of galaxies from the ALMA+MUSE HzRG sample, named in column (1), are based on \ion{He}{ii} $\lambda$1640 and [\ion{C}{I}](1-0) line fitting shown in columns (2) and (3). The literature redshift is shown in column (4) with the cited source provided in column (5). }
  \label{table:redshifts}
  \begin{tabular}{l D{,}{\, \,\pm\, \,}{-10} D{,}{\, \,\pm\, \,}{-10} D{,}{\, \,\pm\, \,}{-10} l}
    & & & &  \\
   \hline \hline
  Galaxy & 
  \mc{\ion{He}{ii} redshift} & 
  \mc{[\ion{C}{I}] redshift} & 
  \mc{Literature redshift} & 
  Reference \\
    & & & &  \\
      \hline
    & & & &  \\
      MRC 0943-242		& 2.9230,0.0001  	& 2.9215,0.003  	& 2.9230,0.0020 	&  \citet{Roettgering1997} \\
      TN J0205+2242  	& 3.5060,0.0003  	& \dots  	& 3.5061,0.0004 	& \citet{deBreuck2001} \\
      TN J0121+1320 	& 3.5190,0.0002  	& 3.5230,0.004  	& 3.5200,0.0007 	& \citet{Nesvadba2007} \\
      4C+03.24    		& \dots     & 3.5828,0.004      & 3.5699, 0.0003 	& \citet{vanOjik1996} \\ 
      4C+19.71   		& \dots	& 3.5892,0.004 	& 3.5935,0.0007 	& \citet{Nesvadba2017} \\
      TN J1338-1942		& 4.0959,0.0005  	& \dots  	& 4.1057,0.0004 	& \citet{Swinbank2015} \\
      4C+04.11 			& 4.5080,0.0002  	& \dots  	& 4.5100,0.0001 	& \citet{Nesvadba2017} \\
    & & & &  \\ 
      \hline
  \end{tabular}
\end{table*}

  \subsection{[\ion{C}{i}] narrow-band imaging}\label{subsection:mom-0}
Locating [\ion{C}{i}] emission in projection is crucial in determining whether the traced gas is associated with the host galaxy or extended CGM. We use the narrow-band images created with the procedure described in Section \ref{section:alma-details}. The maps are spectrally integrated over velocity intervals which display the spatial extent of [\ion{C}{i}] emission by summing the consecutive channels which exceed the zero-flux level. For the three galaxies where [\ion{C}{i}] is not detected, we do not construct a moment-0 map or continuum-subtract because none of the galaxies show any significant continuum emission. The [\ion{C}{i}] contours in Fig. \ref{fig:galaxy-spectra-images} are rather close to the AGN host galaxies detected with {\it Spitzer}, indicating that our faint [\ion{C}{I}] detections are likely to be associated with the host galaxies.

  \subsection{Neutral carbon spectral line fits}\label{subsection:line-fitting}
In this study, we trace molecular gas via the [\ion{C}{i}] lines which we fit using single Gaussian models. Our [\ion{C}{i}] line spectra are extracted from primary-beam corrected image cubes in 2\arcsec~diameter apertures ($\sim$15 kpc at $z=3.5$) which have similar dimensions as the synthesised beams for each observation. We select this aperture size to ensure that (i) the full central part of the host galaxy is included, and (ii) the extracted 1D spectra have sufficient S/N to provide a stable convergence of the fitting algorithm. The sky co-ordinates of the host galaxies are already known from VLA radio continuum detections and \citep{Carilli1997,Pentericci2000} and {\it Spitzer} near-infrared observations of the host galaxies \citep{Seymour2007,deBreuck2010}. 

A high degree of certainty in the host galaxy positions is provided by the sub-arcsecond astrometry in both the MUSE and legacy VLA observations; hence we can extract spectra of the AGN host galaxies using the positions of the radio cores. For this procedure, fixing the locations, however, does not necessarily optimise the S/N as the peak of the [\ion{C}{I}] emission may have a small spatial offset from the AGN host galaxy. This is still acceptable for our purposes because we only aim to determine the [\ion{C}{I}] content inside the AGN host galaxies which can provide us with an indication of the molecular gas available for star-formation. The synthesised beam sizes are 1\arcsec~in radius which at $z=3.5$ corresponds to a projected size of $\sim$15 kpc which sufficiently covers the host galaxy sizes represented by the {\it Spitzer} IRAC $K$-band contours in the narrow-band images.

On the extracted 1D spectra, we fit the lines in a similar manner as that for the single-peaked \ion{He}{II} lines (see Section \ref{section:HeII-fits}). The extracted spectra are shown in the bottom left panels of Fig. \ref{fig:galaxy-spectra-images} with the results of the fit convergence displayed in Table~\ref{table:alma-ci-detected-galaxy-properties}. The \ion{He}{II} and [\ion{C}{I}] line fits are completely independently of one another.

  \subsection{Neutral carbon line luminosity and molecular gas mass}\label{subsection:line-emission-measures}
From the [\ion{C}{I}] spectral line fitting, we obtain [\ion{C}{I}] line flux density measures to infer line luminosities and [\ion{C}{I}]-derived H$_2$ masses for the radio-AGN host galaxies. For non-detections, we estimate $5\sigma$ flux density upper limits of $S_{[\ion{C}{I}]}\rm dV = 5\sigma_{\rm rms} \sqrt{\delta \varv \Delta \varv}$ where $\sigma_{\rm rms}$ is the root-mean-square (RMS) level in the 1D spectrum estimated over a velocity range of -50 to 50 km s$^{-1}.$ The average channel size is denoted by $\delta \varv.$ This velocity width, $\Delta \varv \simeq 100$ km s$^{-1},$ is based on the assumption that emission from the cold gas is traced by line dispersions of the order $\sim$100 km s$^{-1}.$

We calculate the line luminosity of [\ion{C}{I}], $L'_{[\ion{C}{I}]},$ \citep[e.g.][]{SolomonDownesRadfor1992} as,
\begin{equation}\label{eqn:L-prime-CI}
L'_{[\ion{C}{I}]} = 3.25 \times 10^{4}~S_{[\ion{C}{I}]}~{\rm dV}~\nu_{\rm obs}^{-2}~(1+z)^{-3}~D_{\rm L}^{2},
\end{equation} 
to obtain $L'_{[\ion{C}{I}]}$ in K km s$^{-1}$ pc$^2.$ Here, $S_{[\ion{C}{I}]}\rm dV$ is the integrated flux in mJy km s$^{-1}$, $\nu_{\rm obs}$ is the observed frequency of [\ion{C}{I}] in GHz, $D_{\rm L}$ is the luminosity distance of the source in Mpc and $z$ is its redshift. Alternatively, the line luminosity in solar luminosities (L$_\odot$) may be written as,
\begin{equation}
L_{[\ion{C}{I}]}= 1.04\e{-6}~S_{[\ion{C}{I}]}~{\rm dV}~\nu_{\rm rest}~(1+z)^{-1}~D_{\rm L}^{2}
\end{equation}
where, as in equation \ref{eqn:L-prime-CI}, $S_{[\ion{C}{I}]}\rm dV$ is the integrated flux in mJy km s$^{-1}$, $\nu_{\rm rest} = 492.161$ GHz, the rest frequency of the [\ion{C}{I}](1-0) line, and $D_{\rm L}$ is the luminosity distance of the source in Mpc and $z$ is its redshift.

The [\ion{C}{I}] line luminosity provides an inference for molecular gas mass in solar masses (M$_\odot$) via the equation, 
\begin{equation}\label{eqn:H2-mass}
M_{\rm{H}_2,[\ion{C}{I}]} =  \frac{1375.8}{ Q_{10}} ~\frac{D_{\rm L}^2}{(1+z)}\left[\frac{X_{[\ion{C}{I}]}}{10^{-5}} \right]^{-1}~ \left[ \frac{\rm{A}_{10}}{10^{-7} \rm{s}^{-1}} \right]^{-1}~\left[\frac{S_{[\ion{C}{I}]}\rm{dV}}{\rm Jy~km~s^{-1}}\right].
\end{equation}

\noindent In equation \ref{eqn:H2-mass}, the luminosity distance, $D_{\rm L},$ is in Mpc, the Einstein A-coefficient is $\rm A_{10}=7.93\e{-8}$ s$^{-1}.$ We use a  [\ion{C}{i}](1-0) excitation factor of $Q_{10}=0.48$ and a carbon abundance of $X_{[\ion{C}{I}]} = 3.0\e{-5}$ \citep{Weiss2005}. 

Three galaxies in our sample are not detected in [\ion{C}{I}](1-0). For these sources, the 5$\sigma$ upper limits yield flux estimates of $\lesssim 100$ mJy km s$^{-1}$ where we assume a line-width of 100 km s$^{-1}$. To contextualise these non-detections, we recast our sensitivity limits in terms of molecular gas mass limits, assuming (like for our detections) $\rm Q_{10}=0.48$ and $\rm X_{[\ion{C}{I}]}=3\times 10^{-5}.$ This results in upper limits of $M_{\rm H_2,[CI]} < (0.5-1.0)\e{10}$ M$_\odot$. Unfortunately, no CO observations have been reported for these three galaxies, hence we cannot determine whether the [\ion{C}{I}] upper limits are consistent with CO-derived masses in these three sources.

\section{Results and Discussion}\label{section:results-discussion}
  \subsection{Neutral carbon within the host galaxies}\label{section:CI-hosts}
Based on our results, we find that [\ion{C}{i}] emission is fainter than expected compared to recent detections of this line in HzRGs such as the Spiderweb galaxy \citep{Gullberg2016a,Emonts2018} and PKS 0529-54 \citep{Lelli2018,Man2019}. At least three HzRGs in our observations are not detected in [\ion{C}{i}], while the remaining four have $2-3\sigma$ level line detections. Nevertheless, we consider this faint line emission worth reporting because it overlaps both spatially and spectrally with the \ion{He}{ii} emission seen by MUSE which traces the warm ionised gas within the ISM. Additionally, the [\ion{C}{i}] detections spatially overlap with the AGN based on the VLA-detected synchrotron emission labelled in Fig. \ref{fig:galaxy-spectra-images}. Although deeper [\ion{C}{i}] observations are required to accurately constrain the line widths and the spatial extent of cold gas traced by neutral carbon, we already observe [\ion{C}{i}] faint emission as a necessary motivation for future searches of [\ion{C}{i}] within other samples of AGN host galaxies. 

In our results, the galaxy TN J0121+132 is observed with a $2\sigma$ [\ion{C}{i}] that traces the full velocity dispersion of the galaxy. Its observed line width of $565\pm194$ km s$^{-1}$ is consistent with that of the CO(3-2) detection in this galaxy \citep{deBreuck2003}. In the other galaxies from our sample, we observe narrow [\ion{C}{i}] emission up to a minimum width of 40 km s$^{-1}$ (see Table~\ref{table:alma-ci-detected-galaxy-properties}). The four detections include the previously reported narrow [\ion{C}{i}] line in 4C+19.71 \citep{Falkendal2021}. Due to our different extraction aperture, we recover a slightly different line profile from the same dataset, which has an intermediate in line width between TN J0121+1320 and two galaxies, 4C+03.24 and MRC 0943-242, which have line-widths with the same order of magnitude ($\sim$50 km s$^{-1}$). [\ion{C}{i}] velocity dispersions of this width are not commonly observed and should be followed up with deeper observations to trace the full extent of the [\ion{C}{i}] emission line. We note that there is a significant agreement in redshift between [\ion{C}{i}] and \ion{He}{II}, especially in the galaxy MRC 0943-242 (see Fig. \ref{fig:galaxy-spectra-images}). This further justifies the use of [\ion{C}{i}] lines as a proxy for the systemic redshifts of galaxies (see \S\ref{section:HeII-fits} for more details). Overall, we obtain flux density measures of $S_{[\ion{C}{I}]}\rm dV \leq 100$ mJy km s$^{-1}$ and non-detections being reported with upper-limits in the [\ion{C}{i}] flux density. Both [\ion{C}{i}] line constraints and upper limits of as well as the molecular gas masses inferred are shown in Table \ref{table:alma-ci-detected-galaxy-properties}.

\begin{figure*}
  \centering
  \subfloat[]{\includegraphics[width=0.4\textwidth]{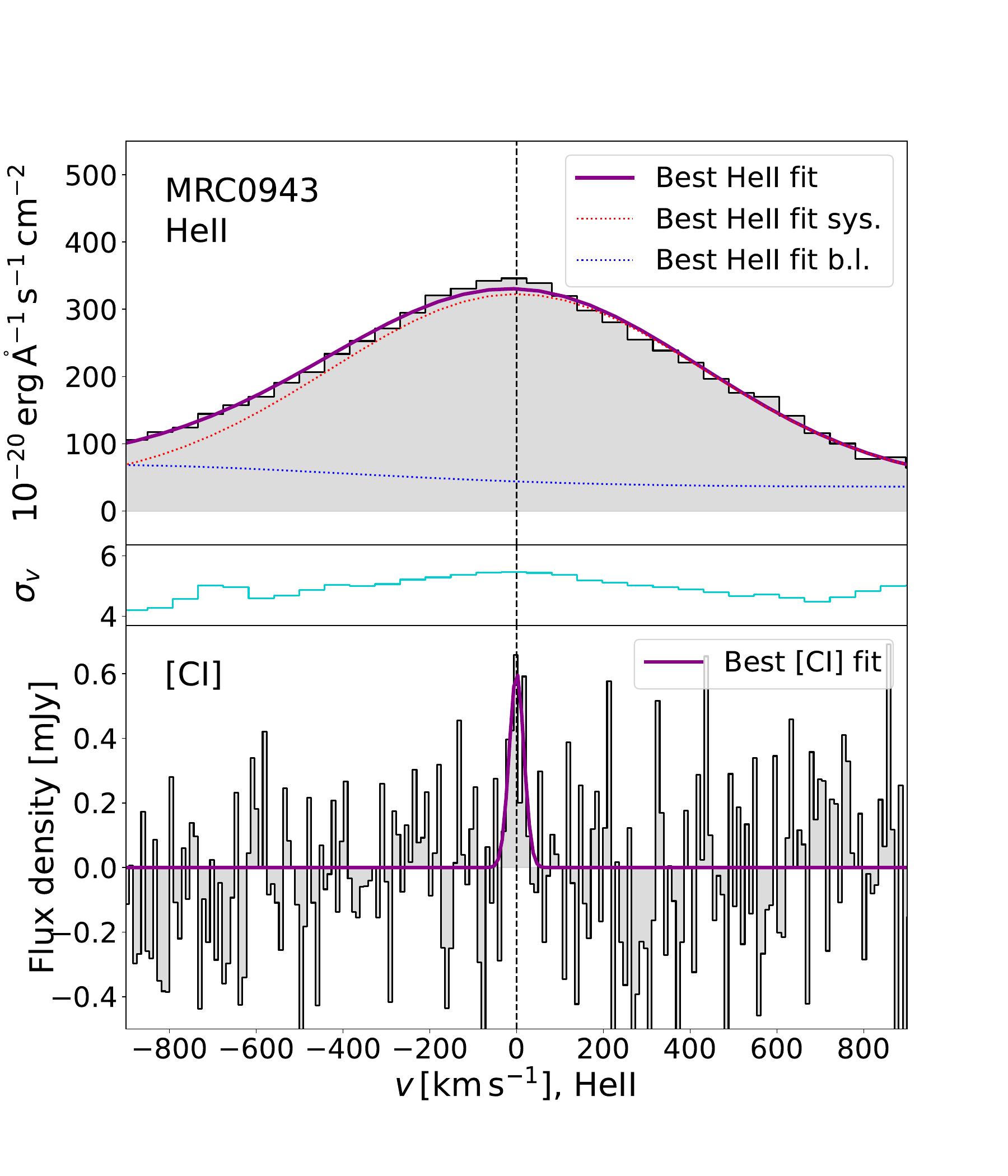}}
  \subfloat[]{\raisebox{5mm}{\includegraphics[width=0.6\textwidth]{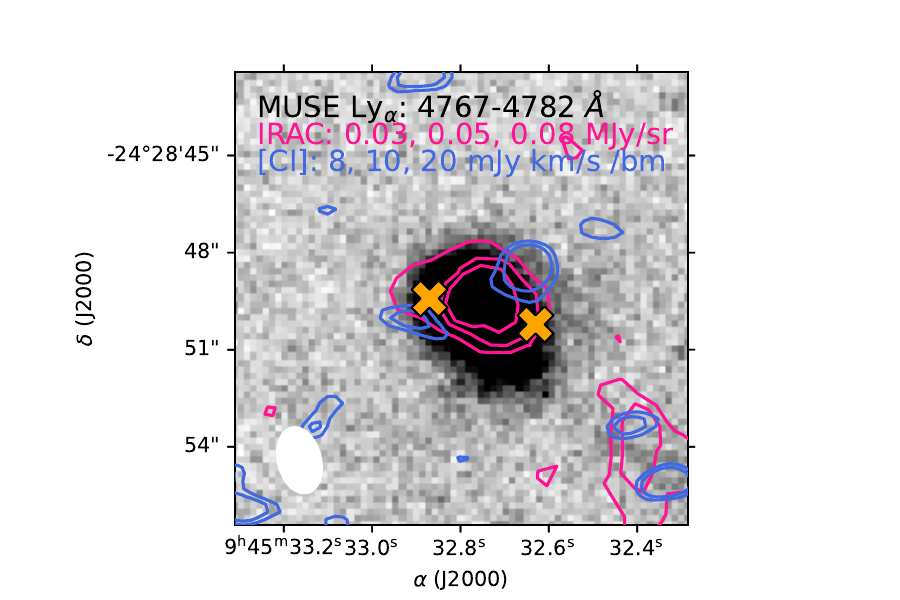}}}\\
      \vspace{-2cm}
  \subfloat[]{\includegraphics[width=0.4\textwidth]{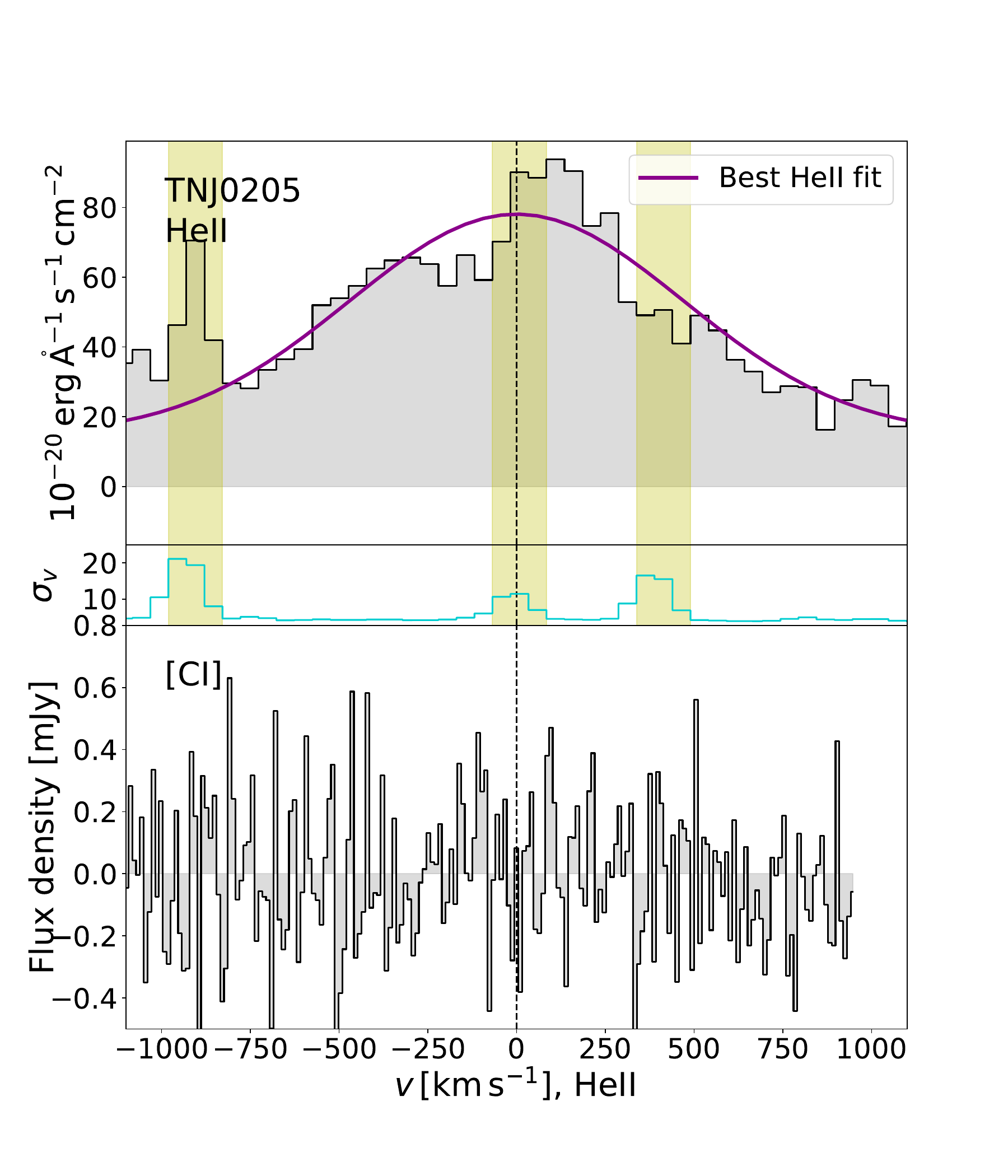}}
  \subfloat[]{\raisebox{5mm}{\includegraphics[width=0.6\textwidth]{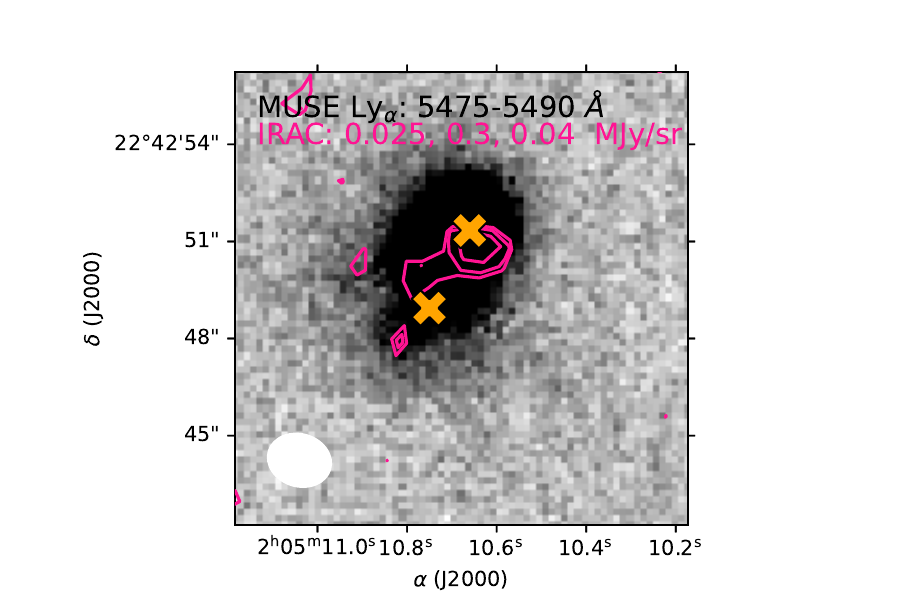}}}\\
  \caption{{\bf Left:} MUSE \ion{He}{ii} and ALMA [\ion{C}{I}](1-0) line spectra of the host galaxies of MRC 0943-42 and TN J0205+2242, in the top and bottom panels, respectively. The $1\sigma$ (rms) noise levels in the [\ion{C}{I}] spectra for MRC 0943-42 and TN J0205+2242 are 0.335 and 0.260 mJy beam$^{-1}$ respectively. The middle panel shows the MUSE spectral variance ($\sigma_v$). All spectra have been extracted from the data cubes with a 1\arcsec\ aperture centred at the peak of the $K$-band continuum. The ALMA [\ion{C}{I}] channel binning is 9 and 11 km s$^{-1}$ for MRC 0943-42 and TN J0205+2242, respectively. When the radio core is not seen, the \ion{He}{ii} is extracted from the peak of the \textit{Spitzer}/IRAC (4.5 $\mu$m) continuum. The yellow vertical bars in the MUSE spectra indicate the locations of skylines with predicted $f_\lambda >$10$^{-16}$ erg s$^{-1}$ $\ang^{-1}$ arcsec$^{-2}$ \citep{Hanuschik2003}. The \ion{He}{ii} line is decomposed into the emission component at the host galaxy (systemic redshift) and the blue-shifted emission is represented by the blue line (b.l.). {\bf Right:} $14 \times 14\arcsec$ MUSE narrow-band image centered on the \Lya line (the covered wavelength range is reported in the image) with \textit{Spitzer} contours overplotted in pink. The [\ion{C}{I}] contours are shown in blue and are in units of mJy beam$^{-1}$ km s$^{-1}$ The orange crosses represent the hotspots of the radio lobes observed by the VLA in its C-band configuration. The contour levels represented for each data set are shown directly in the overlay plots.}
  \label{fig:galaxy-spectra-images}
\end{figure*}

\begin{figure*}
\centering
  \subfloat[]{\includegraphics[width=0.4\textwidth]{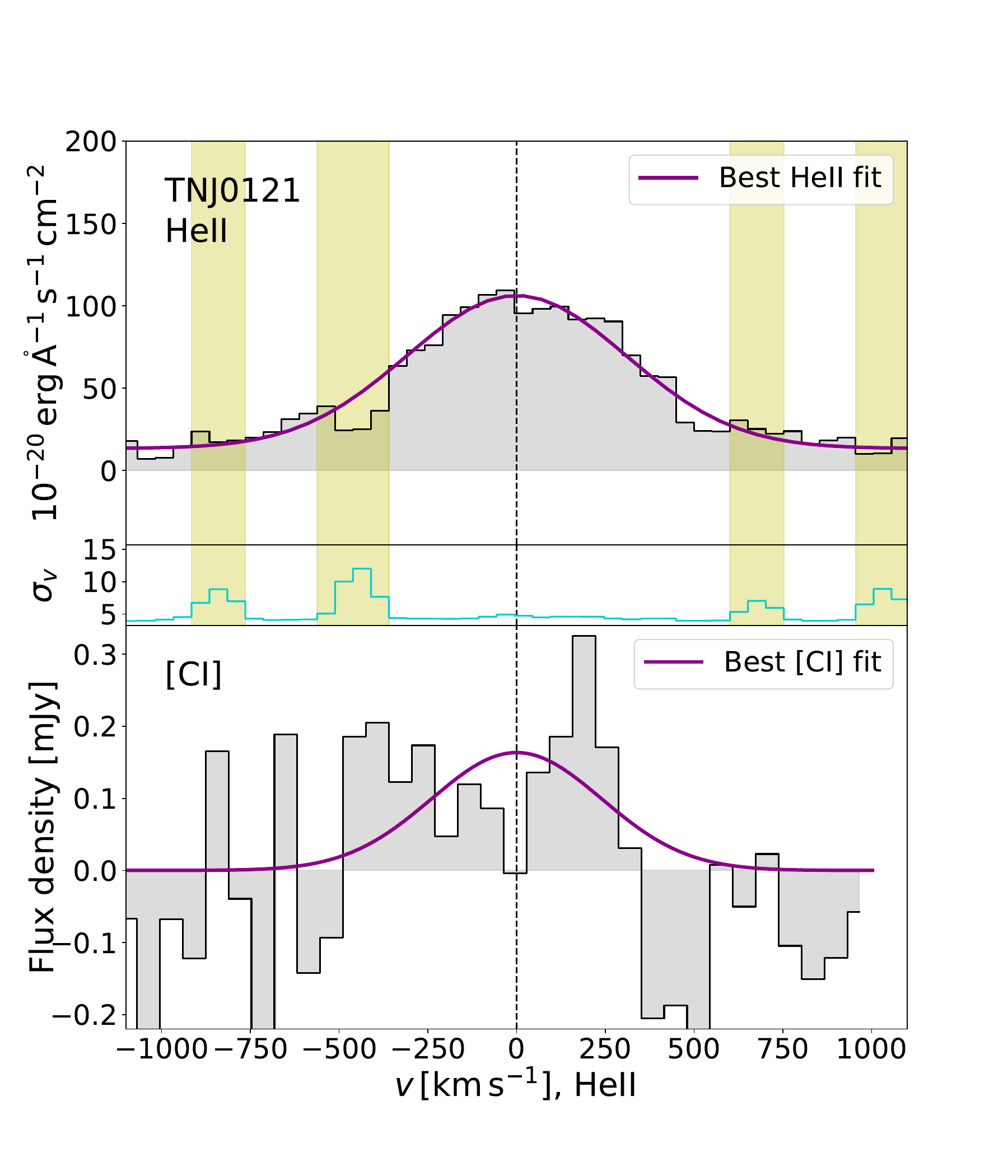}}  \subfloat[]{\raisebox{5mm}{\includegraphics[width=0.6\textwidth]{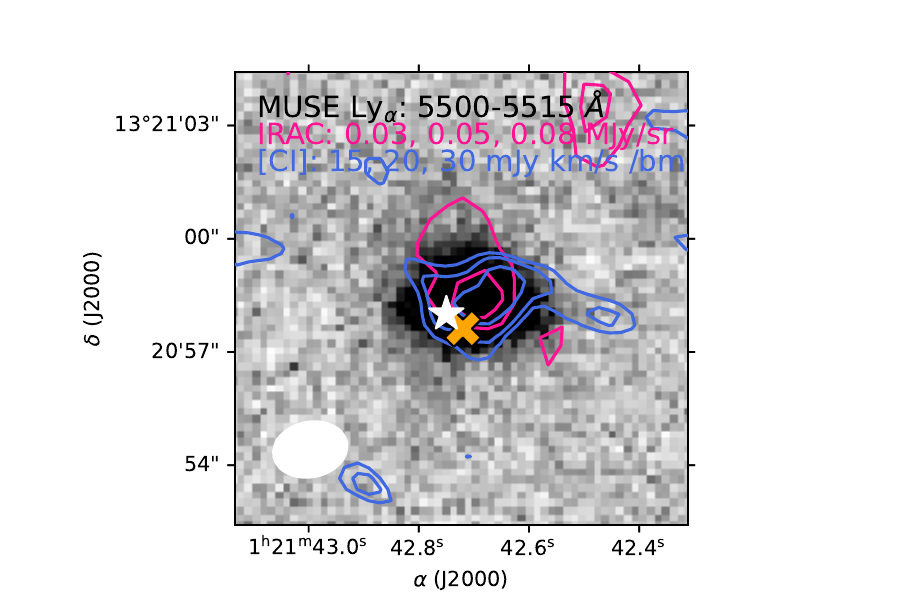}}} \\ 
    \vspace{-2cm}
  \subfloat[]{\includegraphics[width=0.4\textwidth]{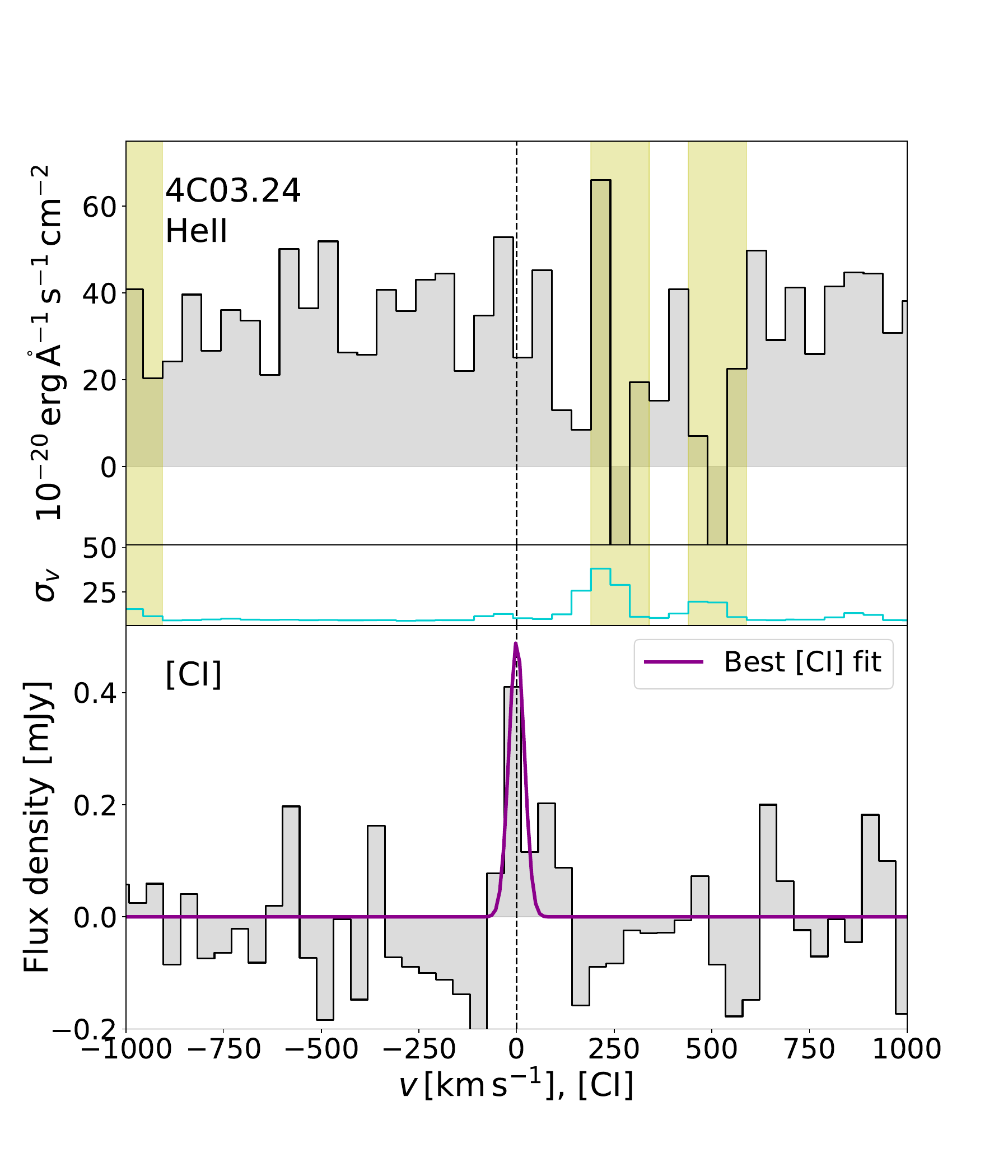}}
  \subfloat[]{\raisebox{5mm}{\includegraphics[width=0.6\textwidth]{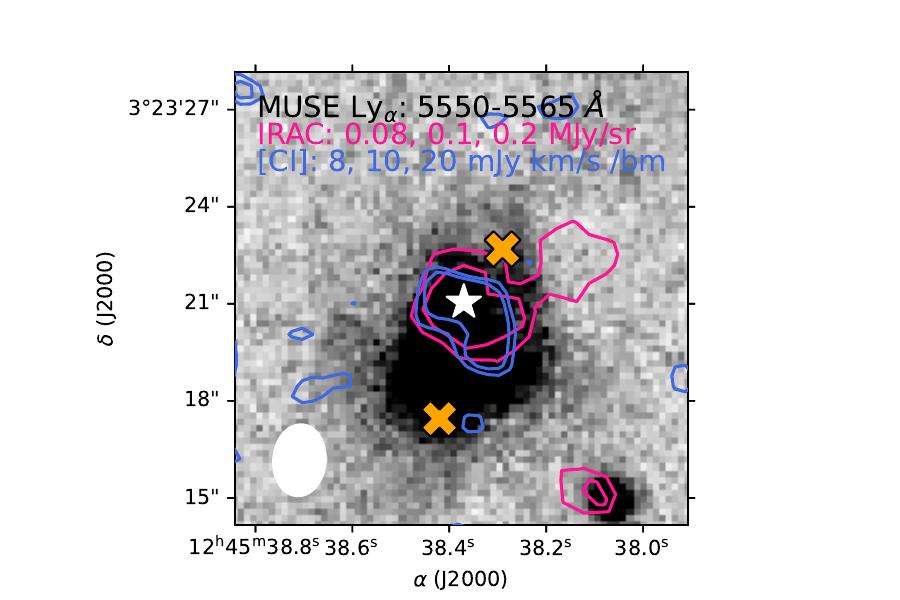}}} \\
\caption*{continued: Here, we show the \ion{He}{ii} and [\ion{C}{I}] spectra alongside the multiwavelength narrow-band images for TN J0121+1320 (top) and 4C+03.24 (bottom). The $1\sigma$ (rms) noise levels in the [\ion{C}{I}] spectra for TN J0121+1320 and 4C+03.24 are 0.311 and 0.189 mJy beam$^{-1}$ respectively. The aperture for [\ion{C}{I}] spectral extraction is centered on the VLA-detected radio core position denoted by the white star. The ALMA [\ion{C}{I}] binning are 64 and 43 km s$^{-1}$ for TN J0121+1320 and 4C+03.24, respectively.}
\end{figure*}

\begin{figure*}
\centering
  \subfloat[]{\includegraphics[width=0.39\textwidth]{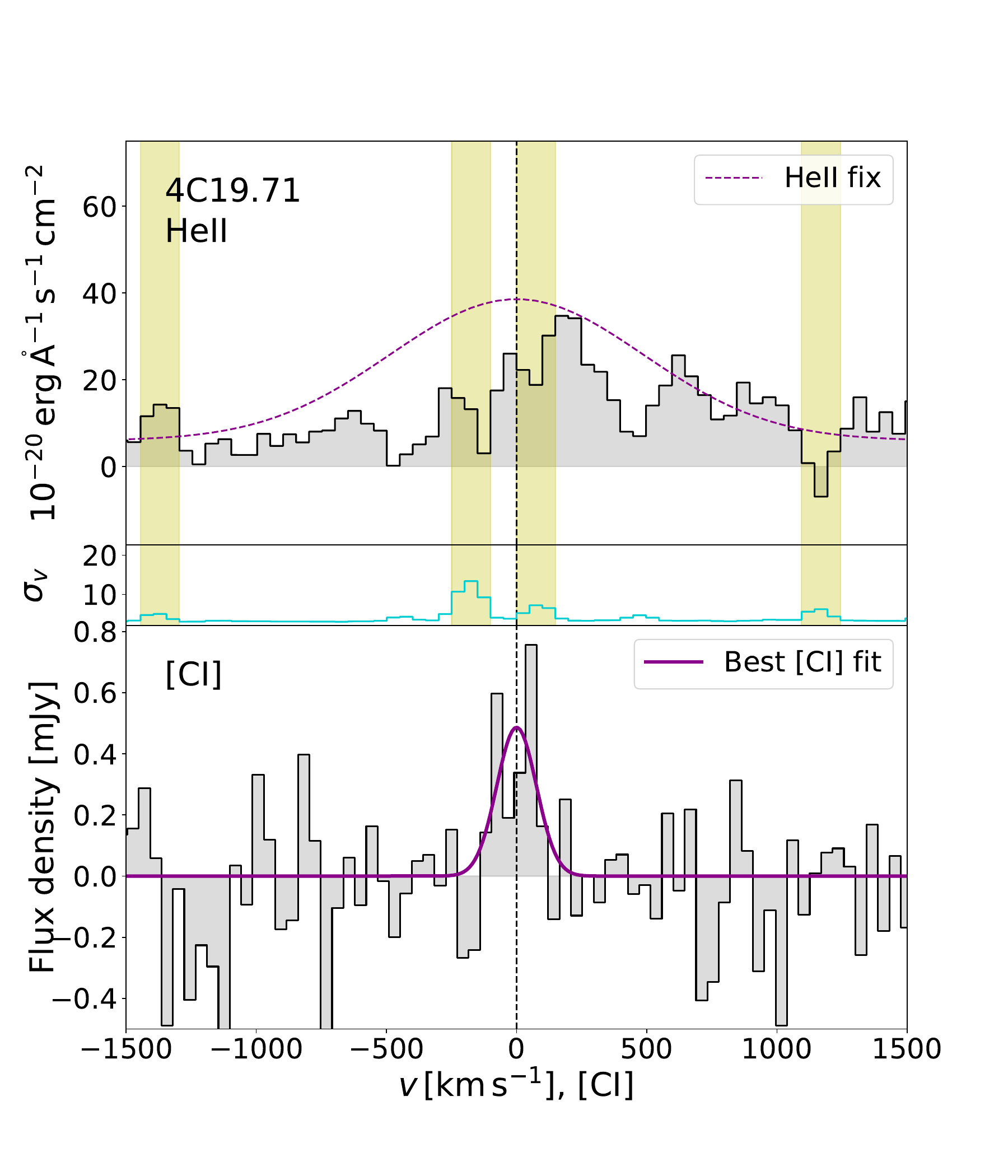}} \subfloat[]{\raisebox{5mm}{\includegraphics[width=0.6\textwidth]{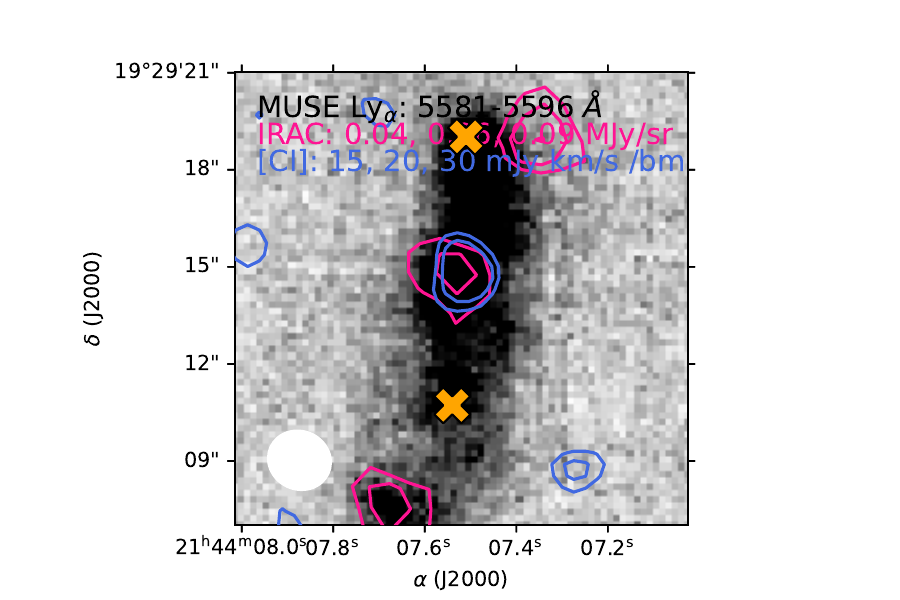}}}\\
  \vspace{-2cm}
  \subfloat[]{\includegraphics[width=0.39\textwidth]{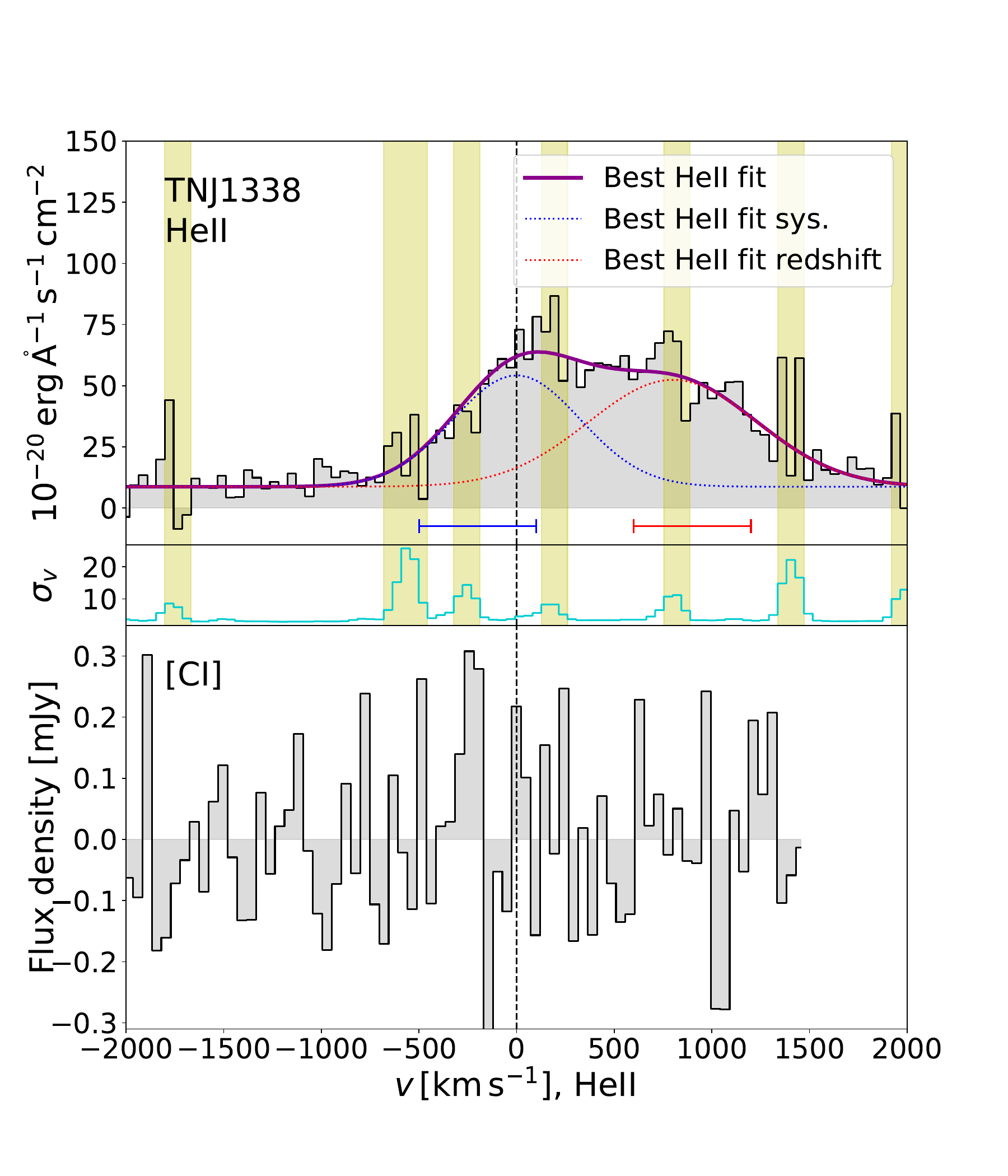}}
  \subfloat[]{\raisebox{5mm}{\includegraphics[width=0.6\textwidth]{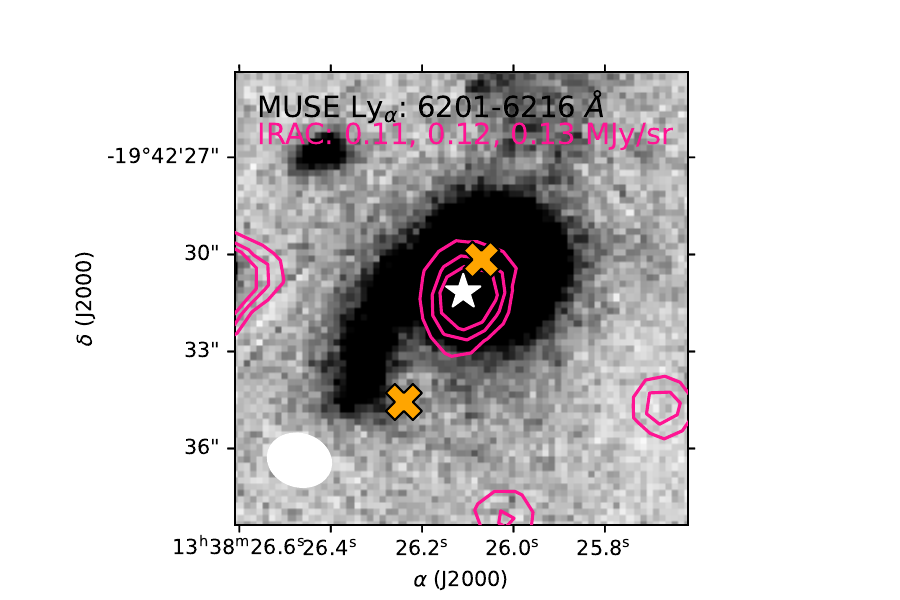}}}\\
    \vspace{-2cm}
  \subfloat[]{\includegraphics[width=0.39\textwidth]{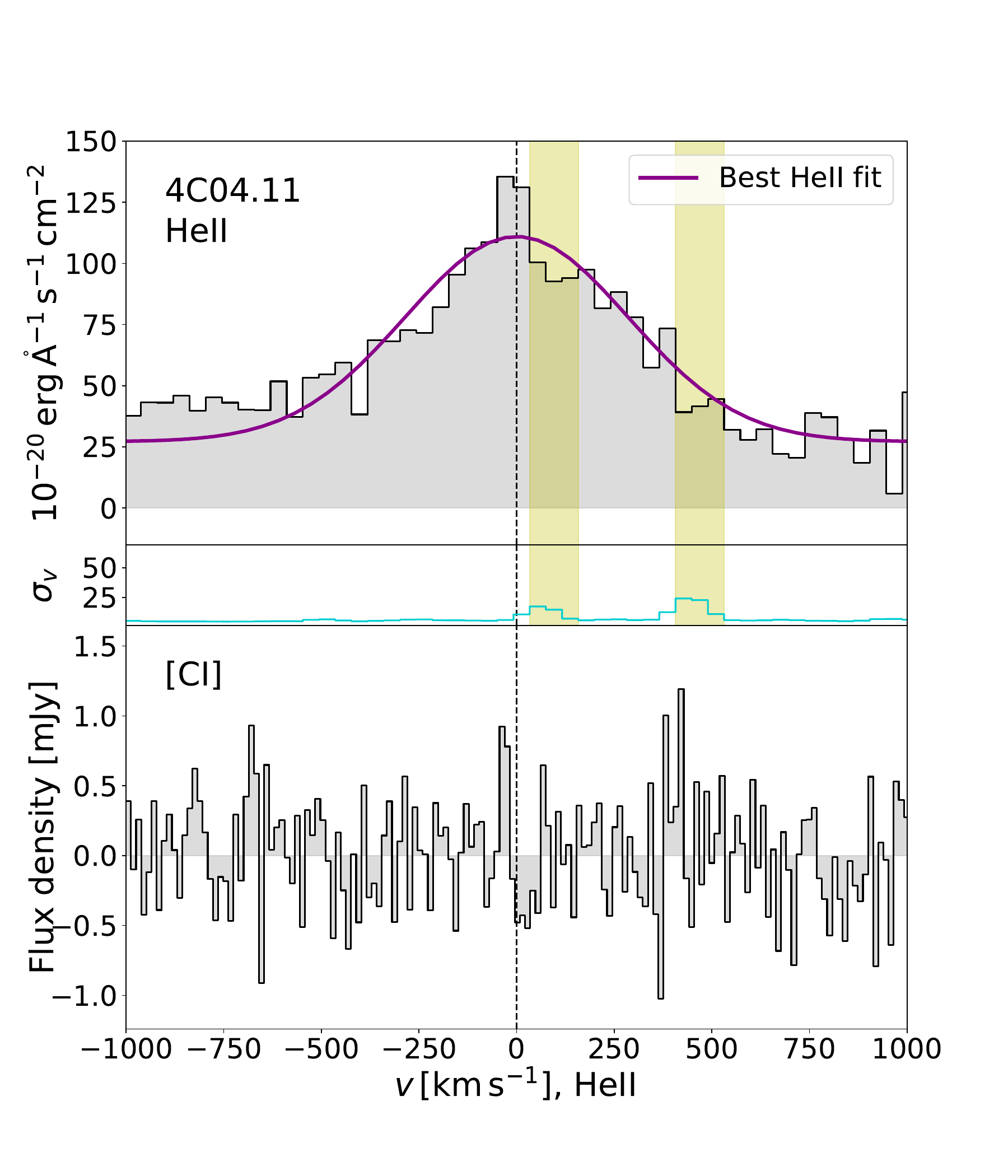}}
  \subfloat[]{\raisebox{5mm}{\includegraphics[width=0.6\textwidth]{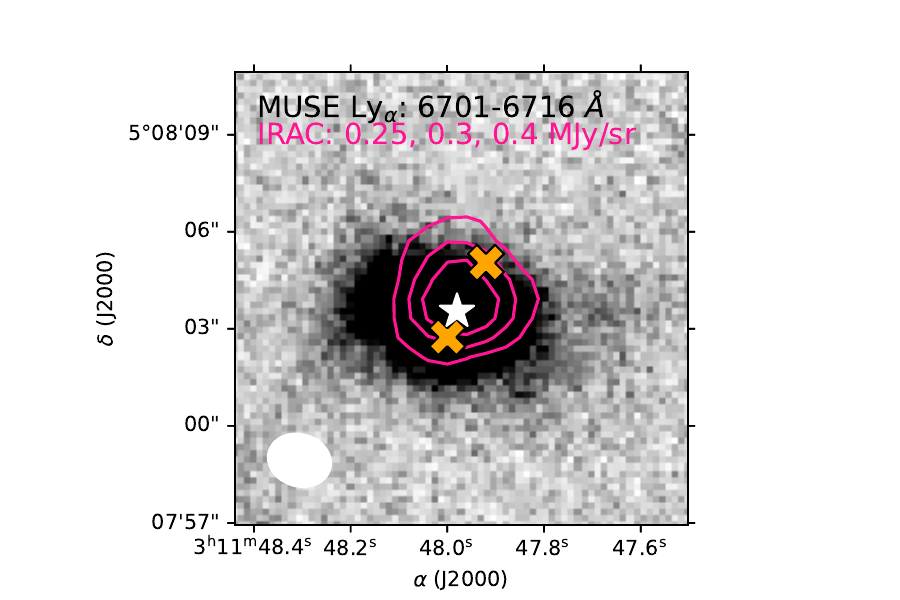}}}\\
     \caption*{continued: Here, we show the \ion{He}{ii} and [\ion{C}{I}] spectra alongside the multiwavelength narrow-band images for 4C+19.71 (top), TN J1338-1934 (middle), and 4C04+11 (bottom). The $1\sigma$ (rms) noise levels in the [\ion{C}{I}] spectra for 4C+19.71, TN J1338-1942 and 4C.04+11 are 0.366, 0.208 and 0.352 mJy beam$^{-1}$ respectively. The ALMA [\ion{C}{I}] spectra are extracted from the $K$-band continuum peak in 4C+19.71 and the VLA-detected radio core TN J1338-1942 and 4C.04+11. The [\ion{C}{I}] binnings are 43, 48 and 13 km s$^{-1}$ for 4C+19.71, TN J1338-1942 and 4C.04+11, respectively.}
\end{figure*}

\begin{table*}
\renewcommand{\arraystretch}{1.2}
  \centering
  \caption[{Observed ALMA [\ion{C}{I}](1-0) line parameters}]{ALMA [\ion{C}{I}](1-0) line parameters. In column (1), we provide the catalog name for the galaxy. Columns (2) to (4) list the observed frequency, the full-width half maximum (FWHM) and integrated flux density of the [\ion{C}{I}] lines, respectively. Column (5) gives the [\ion{C}{I}]-inferred molecular hydrogen gas mass and column (6) provides the star-formation rate reported by \citet{Falkendal2019}; we have obtained SFR uncertainties by propagating the uncertainties in the $\rm L^{SF}_{IR},$ as $\rm \delta SFR = SFR(\delta L^{SF}_{IR}/ L^{SF}_{IR})$. All uncertainties are reported at the $1\sigma$ level. 
    \footnotesize{$^a$ For the 5$\sigma$ non-detections, a line-width of 100 km s$^{-1}$ is assumed.\\
    Stellar masses ($M_\star$) are taken from \citet{deBreuck2010}$^b$ and \citet{Drouart2016}$^c$.}}
  \label{table:alma-ci-detected-galaxy-properties}
  \begin{tabular}{l l D{,}{\,\pm\,}{-2} D{,}{\,\pm\,}{-2} D{,}{\,\pm\,}{-2} D{,}{\,}{0} D{,}{\,}{0} c c D{,}{\,}{0}} 
  \hline \hline
  Galaxy
  & \mc{$\nu_{\rm obs}$}  
  & \mc{FWHM} 
  & \mc{$S_{[\ion{C}{I}]} \rm dV$}
  & \mc{$M_{\rm{H}_2}$} 
  & \mc{SFR} & \mc{$\log{\rm M_\star}$} & $L'_{[\rm \ion{C}{I}]}$ & $L_{[\rm \ion{C}{I}]}$ & \mc{L$^{\rm SF}_{\rm IR}$} \\
& \mc{(GHz)} & \mc{(km s$^{-1}$)} & \mc{(mJy~km~s$^{-1}$)} & \mc{($10^{10}$ M$_\odot$)} & \mc{(M$_\odot$ yr$^{-1}$)} & \mc{(M$_\odot$)} & ($10^8$ K km s$^{-1}$ pc$^2$) & ($10^6$ L$_\odot$) & \mc{($10^{12}$ L$_\odot$)}  \\
  & & & & & & & & & \\
  \hline
MRC 0943-242 & 125.50 & 40,14 & 25,14 & 0.49,0.27 & 41^{+26}_{-33} & 11.22^c & $5.45\pm3.05$ &  2.08 & 0.36^{+0.23}_{-0.29} \\
TN J0205+2242$^a$ & \dots & \dots & <24 & <0.53 & <84 & 10.82^b & $<28.7$ & $<11.0$ & <0.74\\
TN J0121+1320 & 108.81 & 565,194 & 98,55 & 2.60,1.46 & 626^{+267}_{-243} & 11.02^b & $29.0\pm16.2$ & 11.0 & 5.43^{+2.32}_{-2.11} \\
4C+03.24 & 107.39 & 47,16  & 24,8 &  0.65,0.22 & 142^{+240}_{-130} & 11.27^b & $7.28\pm2.43$ & 2.78 & 1.23^{+2.08}_{-1.13} \\
4C+19.71 & 107.24 & 179,60 & 93,32 & 2.54,0.88 & 84^{+173}_{-62} & 11.13^b & $28.3\pm9.74$ & 10.8 & 0.74^{+1.52}_{-0.55} \\
TN J1338-1942$^a$ & \dots & \dots & <93 & <1.08 & 461^{+298}_{-187} & 11.04^b & $<36.8$ & $<14$ & 4.02^{+2.60}_{-1.63}\\
4C+04.11$^a$ & \dots & \dots & <25 & <1.06 & \dots & 11.03^b & $<10.4$ & $<3.98$ & \dots\\
  \hline
  \end{tabular}
\end{table*}

Generally, [\ion{C}{I}] line-widths represent gas kinematics of the cold atomic and molecular gas phase. In our study, the line-widths have been integrated over the full extent of the galactic disk. In MRC 0943-242 and 4C+03.24, the dispersions range from 40-50 km s$^{-1}$ and are insufficiently broad to cover the rotational velocity of the host galaxy and can therefore only trace single molecular clouds within the galaxy. For these two galaxies in our small sample, we observe dispersions that are consistent with galaxy kinematics in 4C+19.17 and TN J0121+1320 which have line-widths of $179\pm60$ km s$^{^-1}$ and $565\pm194$ km s$^{^-1},$ respectively. Such dispersions are comparable to those observed in high-$z$ dusty star-forming galaxies (DSFGs) and SMGs where $\rm FWHM \simeq 200 - 1000$ km s$^{-1}$ \citep{Alaghband-Zadeh2013,Bothwell2017,Nesvadba2019}. In DSFGs and SMGs, X-ray and UV radiation from the AGN and star-forming regions, respectively, heat up gas sufficiently to sustain the broad line dispersions observed in cold gas tracers such as [\ion{C}{I}]. 

Broad CO emission lines have also been detected in low-$z$ ULIRGS which have similar ISM thermodynamic conditions as high-$z$ DSFGs and SMGs where CO gas can exist in gas regions where cosmic rays are dominant \citep{Bradford2003,Papadopoulos2013}. Two galaxies from our sample, TN J0121+1320 and 4C+19.71 have similar line widths as the DSFGs and SMGs. The line widths in these two HzRGs may be tracing the gas kinematics of molecular clouds spread throughout the galaxy. We note, however, that significantly deeper [\ion{C}{I}] observations would be required to obtain reliable kinematic inferences as previously demonstrated in \citet{debreuck2014} where shallow [\ion{C}{II}] data provided a rotation curve based on the cold gas tracers. In fact, these results have been recently fine-tuned using deeper detections in \citet{Lelli2021} as evidence for the importance in sometimes obtaining follow-up observations of low S/N data.

A more direct comparison of our results to previous observations is possible via [\ion{C}{I}] detections in PKS~0529-54 and MRC 1138-262 ($z\simeq2.2$; Spiderweb Galaxy). In PKS 0529-54 ($z\simeq2.6$), the shallow [\ion{C}{I}](2-1) profile was interpreted in two ways: \citet{Man2019} associated both velocity components with distinct star-forming regions within the galaxy, while \citet{Lelli2018} interpreted the observations as evidence for a rotating disk with a circular velocity of 310 km s$^{-1}$. In an upcoming study of the same source, ALMA Cycle 6 observations provide  [\ion{C}{I}](2-1) line emission of $\rm FWHM=615 \pm 56$ km s$^{-1}$ and [\ion{C}{I}](1-0) line emission of $\rm FWHM=848 \pm 230$ km s$^{-1}$ \citep{Huang2023}. Both emission components are broader than the values  reported in Table 4 of \citet{Man2019}. Additionally, the newly detected broadened line emission has made it possible to constrain the redshift from the deeper [\ion{C}{I}] line observations.

For the Spiderweb galaxy, \citet{Gullberg2016b} report a more complex [\ion{C}{I}](2-1) morphology consisting of two spatial components: one of which contains a broad emission lines of width 1100 km s$^{-1}$ and one with a width of 270 km s$^{-1}$. Additionally, \citet{Emonts2018} report a similar velocity profile for the Spiderweb Galaxy in [\ion{C}{I}](1-0) emission seen in observations that unfortunately lack the spatial resolution required to resolve the two spatial components along the spectral axis. In comparison the [\ion{C}{I}](1-0) line detections in 4C+03.24 and MRC 0943-242 from our results have line-widths which are a factor of $5-10$ narrower than what is expected based on the previous studies we have discussed here \citep{Gullberg2016b,Emonts2018,Lelli2018,Man2019}. In the case of MRC 0943-242, however, \citet{Gullberg2016b} report a CO(8-7) FWHM of 43$\pm$13 km s$^{-1}$, fully consistent with the 40$\pm$14 km s$^{-1}$ we find in [\ion{C}{I}].

In terms of [\ion{C}{i}] velocity dispersion, the most anomalous source in our sample is TN J0121+1320 which has a [\ion{C}{I}] line-width of $\sim$600 km s$^{-1}$ and is located above the Main Sequence of star-forming galaxies, as seen in Fig. \ref{fig:SFR-M-star}. Within our sample, this source also has the highest SFR measure (see Table~\ref{table:alma-ci-detected-galaxy-properties}) which places it at the upper reach of the Main Sequence in Fig. 5 from \citet{Falkendal2019} as well. In this work, we have obtained a tentative $2\sigma$ [\ion{C}{I}] detection from which we infer a molecular gas mass of $\sim2.60\e{10}$ M$_\odot$. Furthermore, the [\ion{C}{I}] line dispersion and inferred H$_2$ mass are both consistent with the galaxy's position above the Main Sequence which is evidence that is still has sufficient fuel star-formation at a relatively high level ($\sim$626 M$_\odot$ yr$^{-1}$) compared to the other HzRGs within our sample. For the HzRGs with narrow line emission, [\ion{C}{I}] may be tracing sub-kpc star-forming regions within the galaxy.

Another anomalous case from our sample is TN J1338-1942 which has a relatively high star-formation rate of $\sim$461 M$_\odot$ yr$^{-1}$ but no traceable [\ion{C}{I}] line emission within its host galaxy. Its inferred molecular gas mass is an upper limit of $<1.08\e{10}$ M$_\odot$. For this source, \citet{Falkendal2019} reported a 92 GHz ALMA detection which coincides spatially with the northern radio lobe. The spectral energy distribution (in Fig. A.52 of \cite{Falkendal2019}) predicts equal contributions from synchrotron and thermal dust emission at $\sim$92 GHz. Given the multi-band {\it Herschel} detection, we are certain the the high SFR measure is valid however the continuum data lacks the spatial resolution to determine whether such a high SFR is spatially coincident with the AGN or rather a component within a jet-cloud interacting region at $\sim$10\,kpc north of the galaxy nuclei. Higher spatial resolution thermal dust continuum detections are thus required to properly constrain the star-formation properties of this source.

We estimate the strength of atomic line cooling by calculating the ratio of the [\ion{C}{I}] line luminosity to L$_{\rm IR}$ luminosity ($8 - 1000~\mu$m). For the radio galaxies sampled in this work, these line luminosity ratios range from $(2 - 15)\e{-6}.$ A similarly radio-loud AGN host at $z\simeq2.2$ (MRC 1138-257) has a luminosity ratio of $5.6\e{-6}$ based on L$_{\rm IR}$ (the starburst component) reported in \citet{Seymour2012} and $L'_{[\ion{C}{I}]}$ from \citet{Emonts2018}. Previous works have shown that the $L'_{[\ion{C}{I}]}$/L$_{\rm IR}$ values for lensed SFGs and SMGs (at $z=2-4$) range from $(5 - 20)\e{-6}$ \citep{Nesvadba2019}. For lensed, dusty star-forming galaxies, this range is $(2-18)\e{-6}$ \citep{Bothwell2017}. For unlensed SMGs at $z\sim2.5,$ this value ranges from $(5-30)\e{-6}$ \citep{Alaghband-Zadeh2013}. A sample of Main Sequence galaxies at $z\sim1.2$ show line luminosity ratios of $(4 - 187)\e{-6}$ \citet{Valentino2018}. The $L'_{[\ion{C}{I}]}$/L$_{\rm IR}$ values obtained for the high-$z$ radio galaxies sample in this work, are comparable to those for other high-$z$ galaxy populations but trace the lower end of their $L'_{[\ion{C}{I}]}$/L$_{\rm IR}$ ranges. We summarise all $L'_{[\ion{C}{I}]}$/L$_{\rm IR}$ ratios for our sample as well as those from the literature in Table \ref{table:L_CI_IR_line_ratios}. 

\begin{table}
  \centering
  \caption[]{The [\ion{C}{I}](1-0) to infrared (8 - 1000 $\mu$m) luminosity ratios ($L'_{[\ion{C}{I}]}$/L$_{\rm IR}$) for the radio-loud AGN hosts in this work compared to dusty star-forming galaxies and Main Sequence ($z\sim1.2$) galaxies.}
  \label{table:L_CI_IR_line_ratios}
  \begin{tabular}{l D{,}{\,\e \,}{0} l} 
  \hline \hline
  Source(s)
  & \mc{$L'_{[\ion{C}{I}]}$/L$_{\rm IR}$}
  & Reference\\
  & \mc{($10^{-6}$)} & \\
    & & \\
  \hline
MRC 0943-242 & 5.8 & This work \\
TN J0205+2242 & \dots & " "   \\
TN J0121+1320 & 2.0 & " "  \\
4C+03.24 & 2.3 & " "  \\
4C+19.71 & 15 & " "  \\
TN J1338-1942 & <3.5 & " "  \\
MRC 1138-257 & 8.2 & \citet{Emonts2018}\\
  \hline
DSFGs & 5 - 20 & \citet{Nesvadba2019}\\  
" "   & 2- 18 & \citet{Bothwell2017} \\
" "    & 5- 30 & \citet{Alaghband-Zadeh2013}\\
MS ($z\sim1.2$)  & 4 - 187 & \citet{Valentino2018} \\ 
  \hline
  \end{tabular}
\end{table}
  
\subsection{Comparing CO and [\ion{C}{I}]-derived molecular gas in high-$z$ radio galaxies}

For two sources with both [\ion{C}{I}] and CO detections, we can make a direct comparison between the molecular gas tracers. In MRC 0943-242, \citet{Gullberg2016a} report a CO(8-7) detection and a CO(1-0) upper limit. Scaling the CO(8-7) detections of 0.33 Jy km s$^{-1}$ in {\it Yggdrasil} (the AGN host galaxy) to the 0.54 Jy km s$^{-1}$ in the companion galaxy {\it Loke} which is detected in both CO(1-0) and CO(8-7) and implies an inferred mass of $M_{\rm H_2,CO} \sim 1.4 \e{10}$ M$_\odot$. Such a mass, however, is less likely to be a proper constraint than an upper limit as the CO line emission at {\it Yggdrasil} will likely have a higher excitation level due to the presence of the AGN than that seen at the position of {\it Loke}. Our $M_{\rm H_2,[CI]} = (4.9 \pm 2.7) \e{9}$ M$_\odot$ is therefore consistent with these previously reported CO observations. Similarly, at the host galaxy of TN J0121+1320, the CO(4-3) line flux of 1.2 Jy km s$^{-1}$ from \citet{deBreuck2003} leads to an inferred mass of $M_{\rm H_2, CO(4-3)} \sim 7.0\e{10}$ M$_\odot$, when we assume a CO(4-3)/CO(1-0) ratio of 0.5 and an $\alpha_{\rm CO} = 0.8~\rm{M}_{\odot} / (K~km~s^{-1}~pc^2)$. This is a factor of $\sim3$ larger than the [\ion{C}{I}] inferred mass of $\rm M_{H_2,[\ion{C}{I}]} = (2.60 \pm 1.46)\e{10}$ M$_\odot$. The caveat here is that the CO(4-3) observations from \citet{deBreuck2003} were obtained over a synthesised beam of 8\arcsec$\times$4\arcsec and may therefore include multiple components that are not included in the [\ion{C}{I}] spectrum extracted over the 1\arcsec\ diameter aperture, in our results. As expected, the [\ion{C}{I}] and CO gas estimates do not lead to equivalent results due to the assumptions on $\alpha_{\rm CO}$ and $\rm X_{[\ion{C}{I}]}$ and $\rm Q_{10},$ which are more appropriate for populations of galaxies, but have large uncertainties for individual galaxies. If the the CO-inferred H$_2$ mass were to be made equivalent to that of [\ion{C}{I}], we would require a [\ion{C}{I}] flux density of $S\rm dV_{[\ion{C}{I}]} \geq 264$ mJy km s$^{-1}$ which is inconsistent with the observations presented in this work.

We conclude that the uncertainties on the H$_2$ masses derived from the high-$J$ CO lines are too high to provide a detailed prediction for the detectability of [\ion{C}{I}]. We have, however, with the current depth reached corresponding H$_2$ masses lower than those previously reported in HzRGs \citep[e.g.][]{Papadopoulos2000,deBreuck2003,deBreuck2005,Ivison2008,Emonts2014,Emonts2015,Gullberg2016b}. With sufficiently longer integration times on these sources, targeting the [\ion{C}{I}] line with an interferometer such as ALMA, there is a probability for detection in the sources where we have derived upper limits. 

\subsection{Star-formation efficiency}
The star-formation efficiency (SFE) is defined as $\rm SFE = SFR/M_{\rm gas}$, where $M_{\rm gas}$ should include all gas phases (\ion{H}{I}, \ion{H}{II}, H$_2$). In HzRGs, the H$_2$ component was often assumed to be the dominant one \citep[e.g.][]{deBreuck2003}, though in some cases the neutral \ion{H}{I} gas causing the \Lya can reach masses of order 10$^{10}~\rm M_{\odot}$ \citep{Gullberg2016b,Kolwa2019,Falkendal2019}. Using the lower $H_2$ masses when concentrating on the AGN host galaxies only from our ALMA [\ion{C}{I}] data, we now compare HzRGs with other high redshift galaxies in terms of SFE and gas fraction $f_{\rm gas} = M_{\rm gas}/(M_{\rm gas} + M_\star$).

\begin{figure}
  \centering
  \includegraphics[width=\columnwidth]{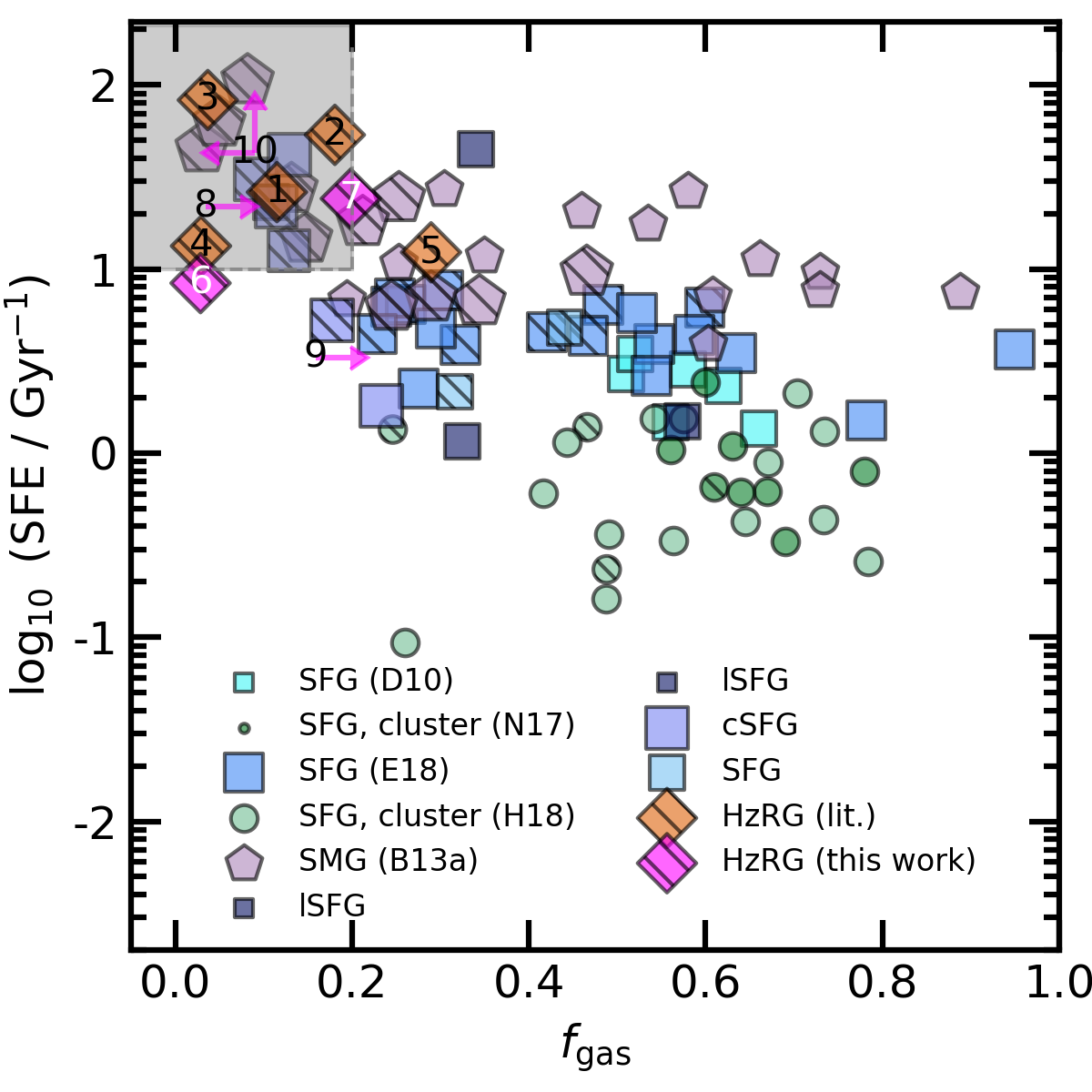}
  \caption[SFE as a function of $f_{\rm gas}$ for HzRGs, SMGs and SFGs]{Star-formation efficiency as a function of molecular gas fraction. The HzRGs from our sample (magenta) and from the literature (orange) are numbered as described in the text. We compare our sub-sample of HzRGs with star-forming galaxies at $z\sim1.4$ \citep[H18;][]{Hayashi2018}, at $z\sim1.6$  \citep[N17;][]{Noble2017}, at $z\sim1.4-1.6$ \citep[D10;][]{Daddi2010} as well as a $z\sim1.2-4.1$ sub-mm galaxy population \citep[B13a;][]{Bothwell2013a}, lensed SFGs (lSFGs - navy blue), compact SFGs (cSFGS - blue) and SFGs (faint blue), see referenced in the text. HzRGs from the literature are numbered as (1) PKS 0529-549 at $z\simeq2.6$  \citep{Man2019}; (2) 4C41.17 at $z\simeq3.8$ \citep{deBreuck2005}; (3) MRC 0152-209 at $z\simeq1.9$ from \citep{Emonts2015}; (4) MRC 1138-262 at $z\simeq2.2$ \citep{Gullberg2016b}; and (5) 4C60.07 at $z\simeq3.8$ \citep{Greve2004}. The hatched symbols represent galaxies in each sample with stellar masses of $M_\star \gtrsim 10^{11}$ M$_\odot.$ The grey box represents the high SFE and low $f_{\rm gas}$ region adopted in previous studies \citep[e.g][]{Man2019}. Details on the tracers used to obtain $f_{\rm gas}$ are provided in the text. The HzRG uncertainties for sources (6) and (8) are propagated from errors in SFR errors based on the work of \citep{Drouart2014} and \citet{Falkendal2019}, $M_\star$ errors from \citet{deBreuck2010} and $M_{\rm H_2}$ errors reported in Table \ref{table:alma-ci-detected-galaxy-properties}.}
  \label{fig:SFE-fgas-plot}
\end{figure}

For our HzRG sample, we use the molecular gas mass (or their limits) as derived from the [\ion{C}{I}](1-0) emission. Due to the lack of systematic measurements, we neglect the \ion{H}{I} component, which may lead to an under-estimate of $M_{\rm gas}$ by a factor up to two. We then use the SFRs from \citet{Falkendal2019} to derive the SFE. To calculate the $f_{\rm gas}$, we consider the radio galaxies' stellar masses from \citet{Seymour2007} and \citet{deBreuck2010}, where the upper limits of 4C+03.24 (= USS1243+036) and 4C+19.71 (= MG2144+1928) are considered detections because their observed $K-$band detections are fully consistent with the {\it Spitzer} 3.6 and 4.5 $\mu$m photometry. We enumerate our HzRGs in Fig. \ref{fig:SFE-fgas-plot} from (6) to (10) where (6) MRC 0943-242, (7) TN J0121+1320, (8) 4C+03.24, (9) 4C+19.71 and (10) TN J1338-1942. Note TN J0205+2242 is not included as both SFR and M$_{\rm H_2}$ are upper limits, while 4C+04.11 does not have an estimate of the stellar mass. Furthermore, we provide the gas fractions and star-forming efficiency in Table \ref{table:HzRG-sample-fgas-SFE} for galaxies within our sample where these values are either limits or constraints. 

We also include SFE and $f_{\rm gas}$ for HzRGs from the literature and number them (1) to (5). These sources are (1) PKS 0529-549 at $z\simeq2.6$ from \citet{Man2019} who use [\ion{C}{I}] to trace H$_2$; (2) 4C41.17 at $z\simeq3.8$ from \citet{deBreuck2005} who use CO(4-3); (3) MRC 0152-209 at $z\simeq1.9$ from \citet{Emonts2015} who use CO(1-0); (4) MRC 1138-262 at $z\simeq2.2$ from \citet{Gullberg2016b} who use [\ion{C}{I}](2-1); and (5) 4C60.07 at $z\simeq3.8$ from \citet{Greve2004}, who use CO(1-0). Fig. \ref{fig:SFE-fgas-plot} also includes a sample of a {\it Herschel}-detected, lensed SFGs (lSFGs) based on their magnification-corrected SFR and $M_\star$ measures \citep{Sharon2013,Bothwell2013b,Dessauges-Zavadsky2015,Nayyeri2017}. In the figure, we have also included six compact SFGs (cSFGs) \citep{Tadaki2015,Spilker2016,Popping2017,Tadaki2017b,Barro2017}. A sample of normal SFGs have been adapted from \citet{Tadaki2015} and are shown in the figure as well.

In Fig. \ref{fig:SFE-fgas-plot}, we compare the HzRG sample with high-redshift compact, normal and lensed star-forming galaxies as well as sub-millimeter galaxies. To derive SFE and $f_{\rm gas}$, various tracers and methods have been used. The star-forming galaxies (SFGs) from \citet{Daddi2010} reside within clusters at $z\sim1.4-1.6$ and have molecular masses derived from CO(2-1). In \citet{Noble2017}, SFGs at $z\sim1.6$ have H$_2$ gas masses derived from CO(2-1). In \citet{Hayashi2018}, cluster-centric SFGs at $z\sim1.46$ have molecular gas masses derived from CO(2-1). The sub-millimeter galaxies (SMGs) in the diagram have molecular gas masses inferred from CO line emission in several transitions from $J_{\rm up}=2-7$ \citep{Bothwell2013a}. Additionally, the lensed SFGs, compact SFGs and SFGs from Fig. \ref{fig:SFR-M-star} are included in the plot where the gas fractions are derived from [\ion{C}{I}], CO or the dust continuum. 

From the SFE-$f_{\rm gas}$ plot (Fig. \ref{fig:SFE-fgas-plot}), we find that HzRGs have generally lower $f_{\rm gas}$ and higher SFE than SFGs and SMGs. The HzRGs with constrained $f_{\rm gas}$ and SFE occupy have $\rm SFE \gtrsim 10$ Gyr$^{-1}$ and $f_{\rm gas} \lesssim 0.3$. SFGs with $M_\star \gtrsim 10^{11}$ M$_\odot$ close to those of HzRGs also occupy this high SFE, low $f_{\rm gas}$ region. This result agrees well with observations of galaxies at $z\simeq 2-4$ where high mass galaxies tend to have lower molecular gas abundances than their low stellar mass counterparts of stellar mass below $10^{11}$ M$_\odot$ as displayed in Figs 5 and 6 of \citet{Tacconi2018}. It is also consistent with the observation that AGN galaxies appear to have lower CO(3-2) gas masses than non-active galaxies \citep{Circosta2021}.

\begin{table}
  \centering
  \caption[]{The gas fractions ($f_{\rm gas}$) and star-formation efficiencies (SFE) of the high-$z$ radio galaxies where the $f_{\rm gas}$ and SFE can be shown as either a limit or a constrained value. The host galaxy 4C04.11 does not have a well-constrained $M_*$ and it is, therefore, not shown. }
  \label{table:HzRG-sample-fgas-SFE}
  \begin{tabular}{l r r} 
  \hline \hline
  Galaxy (Fig. \ref{fig:SFE-fgas-plot} ID)
  & \mc{$f_{\rm gas}$} 
  & SFE\\
  & & (Gyr$^{-1}$) \\
  & & \\
  \hline
MRC 0943-242 (6) & $0.03 \pm 0.02$ &  $8.38^{+8.22}_{-7.09}$\\
TN J0121+1320 (7) & $0.20 \pm 0.11 $ &  $24.1^{+16.5}_{-17.0} $\\
4C+03.24 (8) & $>$0.03 & 21.7\\
4C+19.71 (9) & $>$0.16 &  3.30\\
TN J1338-1942 (10) & $<$0.09 & $>$42.6\\
  \hline
  \end{tabular}
\end{table}

  \subsection{Why molecular gas may be undetected within the ISM of radio galaxies at high-$z$}\label{section:H2-in-hosts}
Fig. \ref{fig:SFE-fgas-plot} demonstrates that high-$z$ radio galaxies in this work and others tend to have significantly lower molecular gas fractions than SMGs and SFGs. Furthermore, [\ion{C}{I}] line emission, a molecular gas tracer, in 4/7 of the galaxies sampled in our work have $2-3\sigma$ detections while in 3/7 galaxies, [\ion{C}{I}] emission is too faint to be detected in our $\sim$45 minute integration times. Additionally, the sampled galaxies devoid of [\ion{C}{I}] emission do not have previous CO detections either leading us to conclude that molecular gas is almost or fully depleted within the ISM of the radio-loud AGN host galaxies. In the sections that follow, we briefly discuss the physical mechanisms that would lead to such observations. 
  
   \subsubsection{High-$z$ radio galaxies in a low star-formation rate phase?} 
 Tentative evidence for a decline in SFR for HzRGs at $1.3 < z < 4.0$ has been provided \citep{Falkendal2019}. In this study, 25 HzRGs are shown against the Main Sequence (M.S.) of star-forming galaxies from \citet{Schreiber2015} and \citet{Santini2017}. We have provided a similar plot in Fig. \ref{fig:SFR-M-star} showing the Main Sequence from \citet{Schreiber2015}. In Fig. \ref{fig:SFR-M-star}, a reasonable majority of the HzRGs sampled sit within 0.3 dex region of scatter around the MS average. Generally, sources located a factor of 10 below the Main Sequence are considered to be in a low star-forming phase \citep{Rodighiero2011}. Furthermore, two of the sampled galaxies (TN J0121+1320 and TN J1338-1942) are vertically offset by $\sim$0.15 dex above the MS while three (MRC 0943-242, 4C+03.24, 4C+19.71) are located $\sim$0.15 dex below it. TN J0205+2242 has a SFR upper limit of $<$84 M$_\odot$ yr$^{-1}$ such that even if it were to eventually receive a constrained measure of SFR, this would be low enough to be place it significantly below the MS average. Due to 4C+04.11 not having a constrained SFR, we have no indication for its position in the MS diagram. Overall, we find that radio galaxies at $1.3 < z < 4.5$ are more likely to exist in a predominantly low star-formation rate phase in their evolution where they may lack sufficient molecular gas to continue fuelling star-formation in the ISM at the SFR predicted by the Main Sequence when a constant SFR is assumed from the Kennicutt-Schmidt law \citep{Kennicutt1998}.

A previous epoch of violent and rapid star-formation may have removed the supply of molecular gas \citep[e.g.][]{Scholtz2023}, leaving only trace amounts, explaining the faint line emission in [\ion{C}{I}] and CO observed within the host galaxies. This idea is supported by the [\ion{C}{I}] line-widths in our sample which are predominantly narrow ($\rm FWHM=40-200$ km s$^{-1}$) with the exception of TN J0121+1320 which has a line-width of $\sim$600 km s$^{-1}$ which is uncharacteristic of galaxies undergoing major starbursts.

   \subsubsection{AGN feedback effects on cold gas}\label{discussion:negative-feedback}
The molecular gas available to fuel star-formation could also have been displaced from the ISM, dissociated or heated as 
radio jets propagate through the gas medium \citep{McNamara2012,Fabian2012}. Previous studies have provided sufficient evidence for mechanical feedback occurring in radio AGN host galaxies \citep{Villar-martin1999,Best2005,Merloni2007,Nesvadba2008,Rosario2010,McNamara2012,Hardcastle2012,Ishibashi2014,Williams2015,Mahony2016,Nesvadba2017,Santoro2020}.
The notion that radio jets from the AGN cause cold gas removal has also been examined by theoretical predictions. Cosmological simulations of AGN feedback have demonstrated that galaxies within the stellar mass range $M_\star \simeq 10^{10} - 10^{11}$ M$_\odot$ undergo a higher level of quenching than their low mass counterparts \citep{Weinberger2017,Nelson2019}. 
In the case of radio AGN specifically, the kinetic feedback may be associated with powerful jets \citep{Dave2020,HardcastleCroston2020,Thomas2020}. 

A causal link between starbursts in radio galaxies and gas-rich mergers has been suggested before \citep{Ivison2012}. In these sources, which are also luminous in the far-infrared, turbulence in the cold gas is induced by an energy injection from either X-ray radiation or mechanical feedback (or both) which result in the subsequent, slow termination of star-formation within the ISM \citep{Papadopoulos2008,Papadopoulos2012}. This result is similar to the merger-associated starbursts in the Spiderweb galaxy, a $z=2.2$ radio-loud AGN, wherein turbulent gas dynamics have been traced via broad [\ion{C}{I}] line emission of width $\rm FWHM \simeq 1100$ km s$^{-1}.$ 
  
It is possible that cold molecular gas has been removed from the galaxy ISM via negative kinetic AGN feedback events within our sample. The gas would be entrained by propagating radio jets and accelerated out of the ISM resulting in the decline in cold molecular gas abundance that would ordinarily be traced via [\ion{C}{i}] or CO line emission. We can briefly approximate the kinetic energy injection produced by radio jets relativistic speeds $\rm \varv/c \sim0.1.$ The kinetic jet powers typically measured for radio-loud AGN are within the range $P_{\rm jet} \simeq 10^{46} - 10^{48}$ erg s$^{-1}$ \citep{Carvalho2002,Bicknell2003}. Detailed calculations have already demonstrated that such relativistic jets produce sufficient power to produce outflows of molecular gas from the ISM of radio galaxies \citep{Nesvadba2010,Nesvadba2021}. 

As stated in Section \ref{section:sample-selection}, our sample comprises jetted radio AGN host galaxies where the kinetic mode of feedback is likely to operate \citep{HeckmanBest2014}. Given the results from \citet{Falkendal2019}, there is clear supporting evidence that a considerable number of HzRGs have low star-formation rates compared to common SFGs and that AGN feedback is responsible for removing molecular gas and shutting off the star formation. Hence, given the high jet powers, relatively low SFR, low gas fractions and high SFE's observed in our sample, mechanical AGN feedback is a possible cause for the removal of molecular gas from the ISM of the radio-loud AGN host galaxies in our sample. 

In summary, due to the observed faintness of the molecular gas as traced via [\ion{C}{I}] in the AGN host galaxies galaxies with high stellar masses and powerful radio jets, we tentatively conclude that AGN feedback may be the cause for the low molecular gas fractions. However, we can not rule out an earlier starburst as the cause of the depletion. Without a well constrained star-formation history of radio-AGN out to their first formation redshifts, we are limited from making strictly conclusive statements. 

\section{Conclusions}
We have presented ALMA bands 3 and 4 observations of [\ion{C}{I}] $^3P_1$ $\rightarrow$ $^3P_0$ for seven radio galaxies (radio-loud AGN hosts) at redshifts $z=2.9 - 4.5$ with the goal of tracing molecular hydrogen via neutral carbon emission. Knowing the locations of the host galaxies, we searched for [\ion{C}{I}] emission at the prescribed co-ordinates. 

In our sample, four galaxies are detected with $2-3\sigma$ level [\ion{C}{I}] line emission. The remaining three galaxies are reported as non-detections with [\ion{C}{I}] flux densities given as $5\sigma$ upper limits. [\ion{C}{I}] line widths within three out the seven sources in the sample range from $\sim$ $40 - 180$ km s$^{-1}$ indicative of emission from  bright molecular gas clouds within the ISM. In one of the galaxies, TN J0121+1320 ($\rm SFR \simeq 626$ M$_\odot$ yr$^{-1}$.), a [\ion{C}{I}] line-width of $\sim$600 km s$^{-1}$ is measured and is indicative of cold gas that is rotationally supported. Overall, we have obtained [\ion{C}{I}] flux densities that provide molecular gas mass inferences where the upper limits are $M_{\rm H_2} < 0.65\e{10}$ and the constraints are in the range, $M_{\rm H_2, [CI]} \lesssim (0.5-3)\e{10}$ M$_\odot.$  

We compare the star-formation efficiencies (SFE) and molecular gas fractions ($f_{\rm gas}$) of our sample to other high-$z$ radio galaxy populations as well as star-forming and sub-mm galaxies (SFGs and SMGs) at $z\simeq2.$ Generally, we find that high-$z$ radio galaxies have $f_{\rm gas} < 0.2 $ and relatively high SFE's of $4 - 45$ Gyr$^{-1}.$ Furthermore, three galaxies in our sample (TN J0121+1320, 4C+03.24, and 4C+19.72) have gas fractions below 10\%.

Based on these results, we consider two physical mechanisms that may explain the faintness of the [\ion{C}{I}](1-0) emission within our sample. The first being that the galaxies have experienced vigorous starburst activity at early epochs in their evolution which has caused a period of star-formation quiescence where molecular gas mass is close to depletion and cannot be traced by either [\ion{C}{I}] or CO. This would be followed by a second mechanism -- the kinetic-mode of feedback that operates in jetted and high-$z$ radio-loud AGN host galaxies, such as the sources in our high-$z$ radio galaxy sample which would eject a significant amount of cold gas from the ISM of the galaxies resulting in the faintness of [\ion{C}{I}] and CO line tracers. While such outflows have been observed in the ionized gas of a sample of $\sim$50 HzRGs \citep{Nesvadba2017}, it is unclear if similar outflows are found in the cold molecular gas. Our current data is too shallow to allow for a proper conclusion on which mechanism is primarily responsible for the low molecular gas fractions we have observed in our sample of radio galaxies.

In future, we aim to conduct a follow-up study of the cold gas within the extended haloes of this sample as well as other $z > 2$ radio galaxies that ties in a greater focus on the mm/sub-mm continuum as well as the tracers of ionised gas observed within the optical window of MUSE. Generally, the warm hot ionised component of the circumgalactic halo gas contributes a higher proportion of a galaxies' baryon budget (see \citet{Tumlinson2017}) in the CGM than cold gas which remains rather elusive in high-redshift galaxies at $z > 2$ as has been demonstrated by this study. Hence, a greater focus should be placed on investigating the warm and hot ionised gas within the baryonic haloes of jetted radio-AGN host galaxies.

\section*{Acknowledgements}
SK acknowledges the financial grants of the National Research Foundation (NRF) and the South African Radio Astronomy Observatory (SARAO; www.sarao.ac.za) whose contribution towards this research is hereby acknowledged (2020). The Inter-University Institute for Data Intensive Astronomy (IDiA; www.idia.ac.za) is thanked for their provision of computing resources utilised in this project (2020-). SK thanks Federico Lelli and Helmut Dannerbauer for providing clarifying remarks on early drafts. Thanks to Ryan Trainor for essential feedback on later versions of this paper. A tremendous thanks to the expert on all things related to carbon in galaxies, Padelis Papadopoulos, who provided invaluable feedback on many iterations of this paper throughout the grueling editing phase. \\
This paper made use of calibrated measurement sets provided by the European ALMA Regional Centre network (Hatziminiaoglou et al. 2015) through the calMS service (Petry et al. 2020).\\
AWSM acknowledges the support of the Natural Sciences and Engineering Research Council of Canada (NSERC) and the Dunlap Fellowship at the Dunlap Institute for Astronomy \& Astrophysics, funded through an endowment established by the David Dunlap family and the University of Toronto. \\
CMH acknowledges funding from a United Kingdom Research and Innovation grant (code: MR/V022830/1). For the purpose of open access, the authors have applied a Creative Commons Attribution (CC-BY) license to any author accepted version arising.\\
This paper is based on observations collected at the European Southern Observatory under ESO programmes 097.B-0323(B), 097.B-0323(C), 096.B-0752(A), 096.B-0752(B), 096.B-0752(C) and 096.B-0752(F).\\
This paper also makes use of the following ALMA observations: ADS/JAO.ALMA\#2015.1.00530.S. ALMA is a partnership of ESO (representing its member states), NSF (USA) and NINS (Japan), together with NRC (Canada), MOST and ASIAA (Taiwan), and KASI (Republic of Korea), in cooperation with the Republic of Chile. The Joint ALMA Observatory is operated by ESO, AUI/NRAO and NAOJ.

\section*{Data Availability}
The raw Atacama Large Millimeter/submillimeter Array (ALMA) data underlying this article are available in the {\it ALMA Science Portal} at \url{https://almascience.eso.org/}, and can be accessed with the project ID: {\tt 2015.1.00530.S}. The Multi-unit Spectroscopic Explorer (MUSE) data are available in the {\it ESO Science Archive Facility} at \url{http://archive.eso.org/cms.html}. The processed MUSE data will be shared on reasonable request to the corresponding author. The {\it Hubble Space Telescope} (HST) data are available in the {\it Hubble Legacy Archive} at \url{https://hla.stsci.edu/}. The {\it Spitzer} Space Telescope (SST) data are available on the {\it SHzRGs Archive} at \url{http://www.eso.org/~cdebreuc/shzrg/}. The Karl G. Jansky Very Large Array (VLA) data is available from the VLA Legacy Data Archive at \url{https://data.nrao.edu/portal/}.



\bibliographystyle{mnras}
\bibliography{Faint_carbon_radio-loud_AGN} 








\bsp	
\label{lastpage}
\end{document}